\documentclass[apj,twocolumn]{openjournal}
\usepackage{graphicx}
\usepackage{xcolor}
\usepackage{amsmath}
\usepackage{amsfonts}
\usepackage{amssymb}
\usepackage{upgreek}
\usepackage{txfonts}
\usepackage{acronym}

\usepackage{hyperref}
\hypersetup{
    unicode, 
    colorlinks=true,
    linkcolor=red,
    citecolor=blue,
    urlcolor=magenta,
}

\newcommand{\orcidauthor}[3]{\author{\href{http://orcid.org/#1}{#2 \openin1 Orcid-ID.png \ifeof1 \else \hskip2pt\includegraphics[width=9pt]{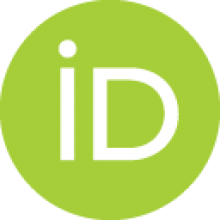}\fi}$^{#3}$}}

\acrodef{NS}[NS]{neutron star}
\acrodef{DM}[DM]{dark matter}
\acrodef{GR}[GR]{General Relativity}
\acrodef{EoS}[EoS]{equation of state}
\acrodef{QCD}[QCD]{quantum chromodynamics}
\acrodef{GW}[GW]{gravitational wave}
\acrodef{ToA}[ToA]{time of arrival}
\acrodef{PPM}[PPM]{Pulse Profile Modelling}
\acrodef{JBO}[JBO]{Jodrell Bank observatory}

\graphicspath{{./}{Figures/}}

\begin{document}

\title{Probing neutron star interiors and the properties of cold ultra-dense matter with the SKAO}


\orcidauthor{0000-0002-4142-7831}{A. Basu}{1}
\orcidauthor{0000-0002-6558-1681}{V. Graber}{2}
\orcidauthor{0000-0001-9208-0009}{M. E. Lower}{3}
\orcidauthor{0000-0002-5470-4308}{M. Antonelli}{4}
\orcidauthor{0000-0003-0746-5364}{D. Antonopoulou}{1}
\orcidauthor{0000-0001-8640-8186}{M. Bagchi}{5,6}
\orcidauthor{0000-0001-6592-6590}{P. Char}{7}
\orcidauthor{0000-0003-1307-9435}{P. C. C. Freire}{8}
\orcidauthor{0000-0002-8255-3519}{B. Haskell}{9,10,11}
\orcidauthor{0000-0002-3407-8071}{H. Hu}{8}
\orcidauthor{0000-0002-0117-7567}{D. I. Jones}{12}
\orcidauthor{0000-0002-3020-9513}{B. Mukhopadhyay}{13}
\orcidauthor{0000-0002-1884-8654}{M. Oertel}{14}
\orcidauthor{0000-0003-2177-6388}{N. Rea}{15,16}
\orcidauthor{0000-0001-5854-1617}{V. Sagun}{12}
\orcidauthor{0000-0002-9581-2452}{B. Shaw}{1}
\author{J.~Singha$^{17}$}
\orcidauthor{0000-0001-9242-7041}{B. W. Stappers}{1}
\orcidauthor{0000-0002-9285-6724}{T. Thongmeearkom}{18,1}
\orcidauthor{0000-0002-1009-2354}{A. L. Watts}{19}
\orcidauthor{0000-0003-2122-4540}{P. Weltevrede}{1}
\author{The SKA Pulsar Science Working Group}-



\email{avishek.basu@manchester.ac.uk}
\email{Vanessa.Graber@rhul.ac.uk}
\email{mlower@swin.edu.au}


\affiliation{$^1$Jodrell Bank Centre for Astrophysics, Department of Physics and Astronomy, The University of Manchester, Manchester M13 9PL, UK}
\affiliation{$^2$Department of Physics, Royal Holloway, University of London, Egham, TW20 0EX, UK}
\affiliation{$^3$ Centre for Astrophysics and Supercomputing, Swinburne University of Technology, PO Box 218, Hawthorn VIC 3122, Australia}
\affiliation{$^4$ CNRS/in2p3, Laboratoire de Physique Corpusculaire (LPC Caen), 14050 Caen, France}
\affiliation{$^5$ The Institute of Mathematical Sciences, Taramani, Chennai 600113, India}
\affiliation{$^6$ Homi Bhabha National Institute, Training School Complex, Anushakti Nagar, Mumbai 400094, India}
\affiliation{$^7$ Departamento de F\'isica Fundamental and IUFFyM, Universidad de Salamanca, Plaza de la Merced S/N, E-37008 Salamanca, Spain}
\affiliation{$^8$ Max-Planck-Institut f\"ur Radioastronomie, auf dem H\"ugel 69, D-53121 Bonn, Germany}
\affiliation{$^9$ Dipartimento di Fisica, Universit\`{a} degli Studi di Milano, Via Celoria 16, 20133, Milano, Italy}
\affiliation{$^{10}$ INFN, Sezione di Milano, Via Celoria 16, 20133, Milano, Italy}
\affiliation{$^{11}$ Nicolaus Copernicus Astronomical Center of the Polish Academy of Sciences, Bartycka 18, 00-716, Warsaw, Poland}
\affiliation{$^{12}$ Mathematical Sciences and STAG Research Centre, University of Southampton, Southampton SO17 1BJ, United Kingdom}
\affiliation{$^{13}$ Department of Physics, Indian Institute of Science, C. V. Raman Road, Bangalore 560012, India}
\affiliation{$^{14}$Observatoire astronomique de Strasbourg, CNRS, Universit\'e de Strasbourg, 11 rue de l'Universit\'e, 67000 Strasbourg, France}
\affiliation{$^{15}$ Institute of Space Sciences (ICE-CSIC), Campus UAB, C/ de Can Magrans s/n, Cerdanyola del Vallès (Barcelona) 08193, Spain}
\affiliation{$^{16}$Institut d'Estudis Espacials de Catalunya (IEEC), Castelldefels, Spain}
\affiliation{$^{17}$High Energy Physics, Cosmology \& Astrophysics Theory (HEPCAT) Group, Department of Mathematics and Applied Mathematics, University of Cape Town, Cape Town 7700, South Africa}
\affiliation{$^{18}$ National Astronomical Research Institute of Thailand, Don Kaeo, Mae Rim, Chiang Mai 50180, Thailand}
\affiliation{$^{19}$Anton Pannekoek Institute for Astronomy, University of Amsterdam, Science Park 904, 1098XH Amsterdam, the Netherlands}


\begin{abstract}
Matter inside neutron stars is compressed to densities several times greater than nuclear saturation density, while maintaining low temperatures and large asymmetries between neutrons and protons. Neutron stars, therefore, provide a unique laboratory for testing physics in environments that cannot be recreated on Earth. To uncover the highly uncertain nature of cold, ultra-dense matter, discovering and monitoring pulsars is essential, and the SKA will play a crucial role in this endeavour. In this paper, we will present the current state-of-the-art in dense matter physics and dense matter superfluidity, and discuss recent advances in measuring global neutron star properties (masses, moments of inertia, and maximum rotation frequencies) as well as non-global observables (pulsar glitches and free precession). We will specifically highlight how radio observations of isolated neutron stars and those in binaries---such as those performed with the SKA in the near future---inform our understanding of ultra-dense physics and address in detail how SKAO's telescopes unprecedented sensitivity, large-scale survey and sub-arraying capabilities will enable novel dense matter constraints. We will also address the potential impact of dark matter and modified gravity models on these constraints and emphasise the role of synergies between the SKA and other facilities, specifically X-ray telescopes and next-generation gravitational wave observatories. 
\end{abstract}




\section{Introduction}

\begin{figure*}
\centering
\includegraphics[width=0.9\linewidth]{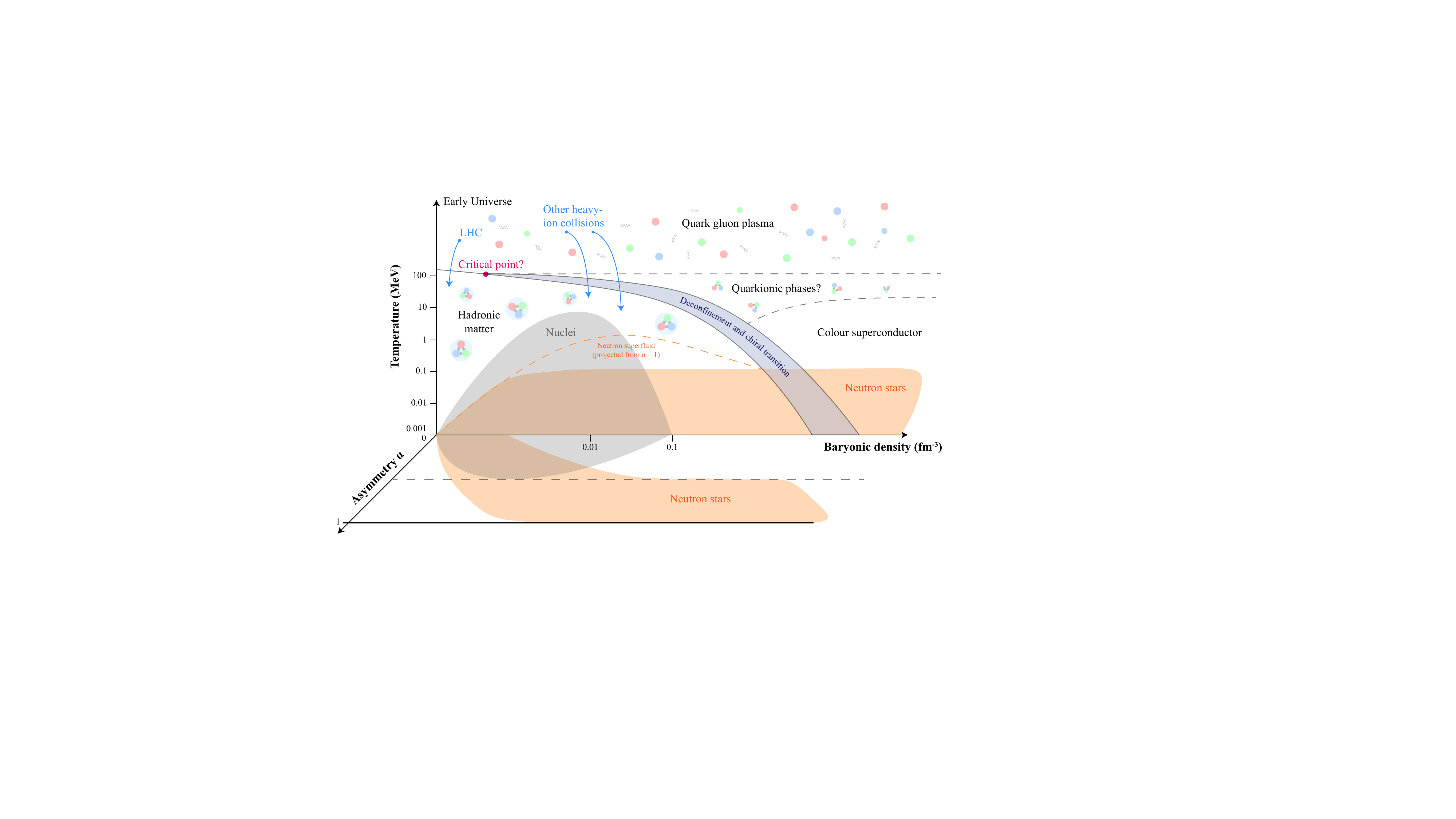}
\caption{The parameter space and states of matter present in \acp{NS}, as compared to terrestrial experiments. The figure shows temperature against baryon number density against asymmetry, $\alpha = 1 - 2 Y_q$, where $Y_q$ is the hadronic charge fraction (generally equal to the ratio of the proton number to the total number of baryons). $\alpha = 0$ for matter with equal numbers of neutrons and protons, and $\alpha = 1$ for pure neutron matter. The orange regions show the parameters space occupied by \acp{NS} projected onto the temperature-baryon density and asymmetry-baryon density planes, respectively. The grey regions show projections onto the same planes for isolated nuclei, which exist up to $\alpha \approx 0.3$. Above this value, one would find a mix of nuclei and light particles.}
\label{fig:trho}
\end{figure*}

Formed in the core-collapse supernovae of massive stars, \acf{NS} interiors are governed by ultra-high densities, low temperatures and large asymmetries in proton/neutron number. Understanding the nature and properties of matter under such extreme conditions is one of the key unsolved challenges in modern science. Because these environments cannot be probed in terrestrial experiments, \acp{NS} are the only laboratories to study matter under high-density, low-temperature and large neutron/proton number asymmetry conditions (see Figure~\ref{fig:trho}) and, thus, advance our general understanding of nuclear physics and \acf{QCD}~\citep{Chatziioannou:2024tjq}. This paper will outline the fundamental role that the SKA will play in this endeavour.

\begin{figure*}
\centering
\includegraphics[width=0.9\linewidth]{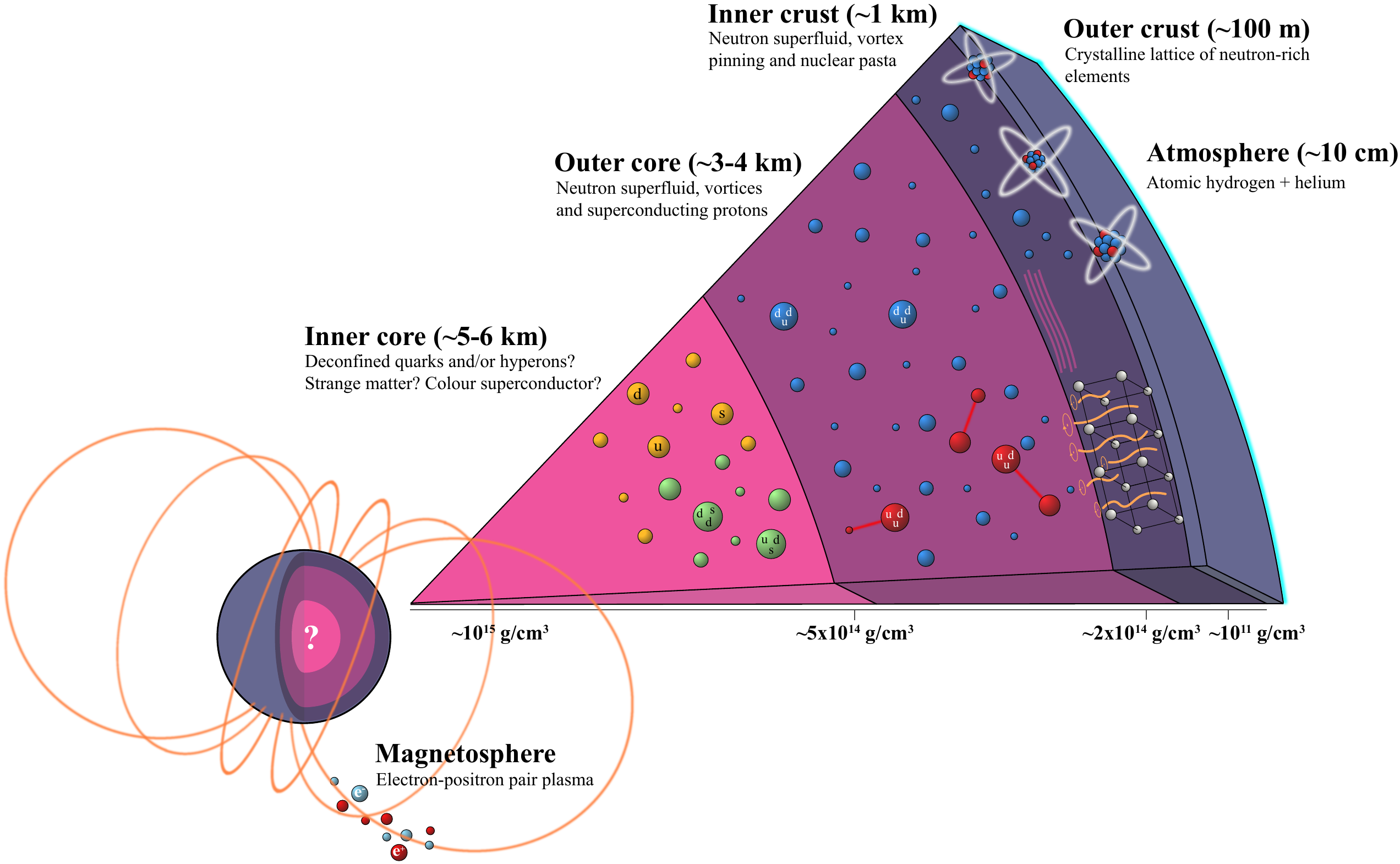}
\caption{Schematic structure of a \ac{NS}: The outer layer---a solid crust of fully ionised nuclei---is supported mainly by electron degeneracy pressure. The inner crust starts around the neutron drip density, $4 \times 10^{11} \, \text{g/cm}^3$, where neutrons begin to leak out of the nuclei. From this point on, neutron degeneracy pressure starts to contribute. At densities of approximately $2 \times 10^{14} \, \text{g/cm}^3$, at the crust-core boundary, nuclei dissolve entirely. In the core, densities can reach up to ten times the nuclear saturation density---the typical density of atomic nuclei---and the pressure due to the repulsive channels of the nuclear interaction is essential to counterbalance gravity.}
\label{fig:schema}
\end{figure*}

A plethora of physical parameters and processes affect the properties of \ac{NS} interiors. Due to their enormous gravitational forces, internal \ac{NS} densities vary by many orders of magnitude resulting in dramatic composition changes across the star (see Figure~\ref{fig:schema}), even though the exact locations of these transitions remain unknown. The primary macroscopic diagnostic to address open questions about the \ac{NS} structure and dense matter interactions is the pressure-density-temperature relation of bulk matter, the \acf{EoS}. Constraining the \ac{EoS} is, thus, essential to inferring key aspects of dense matter microphysics. In addition to uncertainties in composition, we also do not know how ultra-dense \ac{NS} matter behaves dynamically; a question which cannot be answered through \ac{EoS} constraints alone. The key aspect here is that, although \acp{NS} are born hot, they cool down to temperatures well below nuclear energy scales---around $10^9\, {\rm K}$---within months to years~\citep{Page_etal2004}. As a result, \acp{NS} also exhibit the rich phenomenology of low-temperature systems, with at least three distinct macroscopic quantum phases occupying the \ac{NS} interior~\citep{Chamel2017}.

Uncertain composition and dynamical properties are, however, not only of interest from a dense matter perspective. Both play a critical role in astrophysics, with the \ac{EoS} influencing, for example, the dynamics of binary \ac{NS} mergers and the corresponding \acf{GW} signal and nucleosynthesis~\citep{De_etal2018, LIGOScientific:2018hze} as well as core collapse supernovae and the associated neutrino signal~\citep{Janka_etal2007}. Amongst other things, superfluidity (together with the presence of strong magnetic fields) affects \ac{NS} cooling~\citep{Page_etal2004}, the star's rotational evolution in the form of pulsar glitches~\citep{Haskell15,Antonopoulou:2022rpq,
Zhou:2022,Antonelli:2023vpd}, and internal oscillations, particularly relevant for next-generation \ac{GW} facilities~\citep{Andersson:2001kx}. Understanding dense matter is, thus, crucial to understanding numerous high-energy astrophysical phenomena, many of which are multi-messenger events, and SKA radio observations will make essential contributions to this.

The paper is organised as follows. Sections~\ref{sec:unknowns_dmphysics} and~\ref{sec:unknowns_SFphysics} summarise the current state-of-the-art and existing challenges in dense matter physics and dense matter superfluidity, respectively. Section~\ref{sec:radioPSR_NP_connection} gives an outline of how observations of radio pulsars constrain nuclear physics properties, highlighting five distinct classes of \ac{NS} observables and existing nuclear physics constraints. The potential impact of dark matter and modified theories of gravity on these insights is discussed in Section~\ref{sec:DM_modgrav}. We then outline the crucial role that SKA observations will play in constraining dense matter uncertainties in the coming decades in Section~\ref{sec:SKA_expect}. We will specifically discuss the observing modes that are required to unlock SKA's full potential to uncover unknown nuclear physics, while also addressing synergies with other astronomical facilities. Conclusions are presented in Section~\ref{sec:conclusions}.

\subsection{Unknowns in dense matter physics}
\label{sec:unknowns_dmphysics}

The dense matter composing \acp{NS} is governed by the strong interaction in a regime of densities, temperatures, and asymmetries (neutron to proton ratio) neither accessible by terrestrial experiments nor by ab-initio theoretical calculations. Our current understanding of \ac{NS} interiors summarised in Figure~\ref{fig:schema} invokes an outer solid crust with a Coulomb lattice composed of nuclear ions and an electron gas as the outermost region of the star. Moreover, the core is composed of homogeneous strongly interacting matter permeated by an ideal gas of electrons and muons required by electrical charge neutrality.

Only the outer crustal layers contain nuclei for which masses can be measured~\citep{Baym:1971pw,Huang:2021nwk}. Deeper in the crust, descriptions of the properties of neutron-rich nuclei and the neutron fluid close to the crust-core boundary rely on theoretical calculations~\citep{Baym:1971ax,Grill:2014aea,Pearson:2018tkr}. Although model-dependent to some extent, measured properties of finite nuclei and ab-initio approaches to low density neutron matter allow us to sufficiently infer the composition of the inner crust~\citep{Douchin:2001sv,Potekhin:2013qqa}. However, the superfluid and transport properties are more difficult to constrain (see below). For a detailed discussion of the physics of the \ac{NS} crust see, e.g.,~\cite{ChamelHaensel2008} and references therein. When modelling the \ac{NS} core, ab-initio approaches attempt to solve the nuclear many-body problem using few-body interactions as a starting point. While nucleonic two-body interactions at low energies are well constrained by experiments, three- and more-body forces are still a frontier in nuclear physics with some progress at low energies in the last decades due to the development of interactions based on effective field theories~\citep{Tews:2012fj,Hebeler:2013nza,Lynn:2015jua,Drischler:2017wtt,Huth:2021bsp}. Additional uncertainties arise from solving the many-body problem based on different methods, which all include approximations. More phenomenological approaches are based on energy density functionals with parameters determined by nuclear data, benchmark ab-initio calculations, or astrophysical data.

At the high densities reached in \ac{NS} cores, there might be transitions to non-nucleonic states of matter~\citep{Glendenning:1998ag,HeiselbergHjorth-Jensen2000}. Possibilities include the formation of hyperons (strange baryons), $\Delta$-baryons, pion or kaon condensates~\citep{Tolos:2020aln}, or quark matter (leading to the formation of so-called \textit{hybrid} stars), possibly in colour-superconducting states~\citep{Alford:2007xm}. Under some conditions, such hybrid stars exhibit a twin phenomenon: two stars that have approximately the same mass but significantly different radii~\citep{Glendenning:1998ag, Blaschke:2019tbh}. It is even possible that the entire star converts into a lower energy self-bound state consisting of up, down and strange quarks, known as a strange quark star~\citep{Witten:1984rs,Weber:2004kj}. The densities at which such phases would appear, and the degree to which they might co-exist with other phases, are highly uncertain~\citep{Buballa:2014jta,Annala:2019puf, Constantinou:2023ged,Essick:2023fso}. For the non-nucleonic degrees of freedom in the \ac{NS} core, experimental information is scarce or non-existent and uncertainties are much larger than for purely nuclear systems. For reviews on different approaches to describe dense nuclear and \ac{NS} matter see, e.g.,~\citet{Oertel:2016bki,Burgio:2021vgk}. 

Existing insights into the \ac{NS} \ac{EoS} comprise theoretical benchmark calculations and, in particular, chiral effective field theory calculations for low-density neutron matter up to around $1.5$ the nuclear saturation density, $\rho_0$~\citep{Keller:2022crb}, and perturbative QCD calculations at very high densities $(\gtrsim 40 \rho_0)$~\citep{Gorda:2022jvk}. Experimentally, \ac{EoS} constraints arise from data on nuclear masses, nuclear resonances, polarisabilities, heavy-ion collisions, and neutron skin thickness measurements with a recent update by PREX-II and CREX~\citep{PREX:2021umo,CREX:2022kgg}.  \cite{MUSES:2023hyz} present a recent summary of these constraints, but see also \cite{Lattimer:2012xj,Oertel:2016bki} and references therein. Moreover, nuclear physics theory and experiments mostly concern low densities either close to symmetric or pure neutron matter, while \ac{NS} observations probe the neutron-rich (hence highly asymmetric) and high-density matter in \ac{NS} cores. 

From a modelling perspective, Bayesian analysis techniques have gained popularity as a systematic approach to infer the \ac{EoS} from available data. These are either based on a completely uninformed parametric~\citep{Read:2008iy,Lindblom:2012zi} or non-parametric~\citep{Landry:2018prl} representation of the \ac{EoS}, or on nuclear metamodelling techniques~\citep{Margueron:2017eqc,Char:2023fue,Scurto:2024ekq}. Although the crust does not have a significant impact on global \ac{NS} properties like mass $M$, radius $R$ or moment of inertia $I$ except for very low-mass \acp{NS}, it is important to note that only the construction of a so-called `unified' \ac{EoS}\footnote{A unified \ac{NS} \ac{EoS} requires a consistent determination of the core and crust \ac{EoS} and a thermodynamically consistent crust-core transition from the same underlying nuclear model.} ensures quantitatively reliable predictions for these global \ac{NS} properties~\citep{Fortin:2016hny,Suleiman:2021hre} and, therefore, a correct inference of \ac{EoS} properties from data. Moreover, the crust properties are important for interpreting other observables such as pulsar glitches (see below) and magnetar quasi-periodic oscillations~\citep[e.g.,][]{Gabler2018}. Publicly available tools such as CUTER~\citep{Davis:2024nda} allow us to complement any available core \ac{EoS} with a consistent crust.

Future astrophysical data will precisely determine the \ac{NS} \ac{EoS}. However, this may not be sufficient to unambiguously ascertain the composition of matter at the centre of the most massive \acp{NS} in the absence of a phase transition~\citep{Mondal:2021vzt,Xie:2020tdo,Iacovelli:2023nbv,Imam:2023ngm,Char:2025zdy}. This is due to a degeneracy, which arises from different compositions or nuclear interactions leading to the same $\beta$-equilibrated \ac{EoS}. Additional information, e.g., from \ac{NS} phenomena sensitive to transport properties such as cooling rates, pulsar glitches, or oscillation modes, will probably be necessary to fully explore the \ac{NS} interior structure.

We finally note that any inference of \ac{NS} properties from observational data is based on an underlying theory of gravity. While \ac{GR} is typically assumed to hold, modified theories of gravity affect the \ac{NS} structure and induce degeneracies when inferring dense matter properties~\citep{Danchev:2020zwn}. The same holds for the presence of particles beyond the standard model, such as dark matter candidates within \acp{NS}~\citep{Sandin:2008db, Dengler:2021qcq, Sagun:2022ezx}. To fully benefit from upcoming astrophysical data in informing dense subatomic physics, further research is required to reliably disentangle the role of such effects. We briefly return to these issues in Section~\ref{sec:DM_modgrav}.


\subsection{Unknowns in dense matter superfluidity}
\label{sec:unknowns_SFphysics}

Besides their extreme densities, the gravitational confinement renders \acp{NS} sufficiently long-lived for a weak $\beta$-equilibrium to be achieved. As a result, these compact objects can be considered cold, occupying the unique parameter space shown in Figure~\ref{fig:trho}. In fact, \acp{NS} older than a few hundred years are sufficiently cold for the nucleons to form Cooper pairs (analogous to terrestrial electronic superconductors), resulting in the appearance of distinct macroscopic quantum states~\citep{Baym1969, Sauls1989}. In particular, the free neutrons surrounding the lattice nuclei in the inner crust are superfluid, as is the neutron component in the outer core. The latter coexists with a condensate of superconducting protons. Moreover, additional superfluid phases, such as colour superconducting quark phases, might exist in the inner core~\citep{Alford:2007xm, Sedrakian:2018ydt}. 

Superfluid components alter the long-term evolution and dynamics of \acp{NS}. For example, superfluidity affects the star's thermal properties by suppressing nuclear reactions that cool it. Superfluidity also reduces the heat capacity but at the same time opens up new channels for neutrino emission, which can lead to faster cooling~\citep{Page_etal2004}. The corresponding net enhancement of cooling has been invoked to explain the behaviour of the young \ac{NS} in the Cassiopeia A supernova remnant, whose surface temperature may be decreasing faster than expected in standard cooling models due to the recent onset of neutron superfluidity~\citep{Shternin2011, Yakovlev2011}. Even though underlying temperature measurements are challenging~\citep{Posselt2022}, a confirmation of this anomaly would constrain the critical temperature for core superfluidity. Such observations would also place qualitative constraints on the superconducting transition temperature, as enhanced cooling requires this quantum phase to form well before the core superfluid~\citep{Shternin2021}. 

Such constraints are very valuable as transition temperatures directly relate to the highly uncertain superfluid pairing gaps (half the energy required to separate a Cooper pair)~\citep{Baldo1992}. The gaps are calculated assuming that the crustal neutrons and core protons pair in a state with zero spin and angular momentum (isotropic spin-singlet or $s$-wave pairing), while the core neutrons form Cooper pairs of non-zero spin and angular momentum (anisotropic spin-triplet or $p$-wave pairing)~\citep{Gezerlis2014}. However, calculations are complicated by in-medium effects, unknown many-body interactions above saturation, and anisotropy for the core neutrons. Moreover, superfluid energy gaps are typically calculated based on microscopic forces that differ from those considered for the composition and, subsequently, the \ac{EoS}, leading to inconsistencies in modelling superfluid properties. However, chiral effective field theory has recently enabled more consistent approaches for the singlet paired condensates~\citep{Lim2021}.

Superfluidity also affects the stellar dynamics, because---very much as in terrestrial low-temperature experiments---superfluid components can flow relative to the `normal', non-superfluid components, increasing the system's degrees of freedom~\citep{Glampedakis2011}. Remarkably, these microscopic properties lead to observable signatures, e.g., sudden spin-up glitches seen in many young \acp{NS}~\citep{espinoza2011, Basu:2021pyd} and timing noise observed across the entire pulsar population~\citep{Hobbs2010}. Crucially, neutron superfluids rotate by forming tiny quantum vortices, whose interactions with their surroundings control the stellar rotation~\citep{Pines1980, Andersson2006}. The canonical glitch model assumes that these vortices become pinned to the crustal lattice, decoupling part of the stellar interior and leading to the build-up of an angular momentum reservoir~\citep{Anderson1975, Pizzochero2011}. The avalanche-like unpinning of millions of vortices subsequently releases the stored angular momentum by rapidly recoupling the two components and causing the glitch. The shape of this signature, generally described in a body-averaged multi-component picture~\citep{haskell2015IJMPD}, is ultimately determined by the vortex friction, the mechanisms for unpinning and repinning of vortices, and the underlying nuclear properties. 

In the inner crust, friction is primarily driven by the excitation of Kelvin waves along the vortex lines~\citep{Epstein1992, Jones1992, Graber2018ApJ}, while core coupling is dominated by the scattering of electrons off of the neutron vortex magnetic fields~\citep{Alpar1984b, Andersson2006}. The vortex magnetisation relies on entrainment, the non-dissipative coupling between nucleons that controls how mobile the particle species are~\citep{Andersson2006}. Entrainment is also present in the crust due to Bragg scattering off of the lattice nuclei~\citep{Chamel2012}. Although the strength of this process is highly uncertain~\citep{Martin2016, Sauls2020}, it critically affects the angular momentum available for glitches ~\citep{andersson2012PhRvL, chamel2013PhRvL}. Moreover, magnetised neutron vortices might interact with proton fluxtubes in the core, potentially linking the star's rotational and magnetic evolution~\citep{Ruderman1998, Sidery2009}. Such a dissipative process, which could be sufficiently strong to lead to vortex-fluxtube pinning, would also cause strong damping of slow precession~\citep{link_06}. The uncertain nature of the proton superconductor further complicates the situation. While magnetic flux is generally assumed to be confined to a regular array of quantised fluxtubes, flux might be inhomogeneously distributed in the \ac{NS} core due to the strong coupling between the protons and neutrons~\citep{Wood2022}. Our understanding of the stellar magnetism could be further complicated by colour-superconducting phases in the \ac{NS} centre, which might also carry fluxtubes~\citep{Alford2010, Haber2018}. Finally, note that additional complexity arises because vortices (and fluxtubes) might not be straight but could instead form tangles~\citep{Andersson2007}. The impact of such superfluid turbulence on the large-scale stellar dynamics remains poorly understood~\citep{haskell20}. 

While theoretical progress in understanding the complicated processes outlined above remains essential, designated observing campaigns with the SKA will provide an unparalleled view into dense matter properties beyond standard \ac{EoS} constraints~\citep{HaskellSedrakian}. As discussed below, high-precision timing with the SKAO's telescopes offers a unique way to probe related physics, surpassing existing constraints on the superfluid energy gaps from cooling observations and rough estimates of the superfluid moment of inertia from glitch observations.


\section{Connecting pulsar observations to dense matter physics: state-of-the-art}
\label{sec:radioPSR_NP_connection}

\subsection{General concepts}
\label{sec:general_concepts}

The Tolman-Oppenheimer-Volkoff (TOV) equations~\citep{Tolman1939, Oppenheimer1939} that control the \ac{NS} structure provide a unique correspondence between macroscopic \ac{NS} observables, such as mass, radius, or moment of inertia, and the dense matter \ac{EoS}. In the zero-temperature limit, which is generally sufficient to describe the bulk structure of \acp{NS}, the \ac{EoS} relates the \ac{NS} density with the pressure, dependent on the underlying microphysics and nuclear interactions. Corresponding uncertainties outlined in Section~\ref{sec:unknowns_dmphysics} lead to numerous \acp{EoS}, which in turn can be directly mapped onto the $M$-$R$ plane~\citep{Lindblom1992} or $M$-$I$ plane~\citep{Lattimer2005} for \acp{NS} (see Figure~\ref{fig:EOS_mass-radius}). Measurements of, or constraints on, \ac{NS} observables thus allow us to test the viability of different \ac{EoS} models and the stellar composition. While direct measurements of \ac{NS} radii in the radio are unfeasible, and indirect access through $I$ measurements is highly complex (see below), \ac{NS} rotation frequencies and masses are readily available. 

The key technique which underlies these measurements is high-precision radio pulsar timing~\citep{Hobbs2006}. Pulsar timing is based on meticulously tracking every rotation of a \ac{NS} over long time periods---sometimes for decades. Comparing the corresponding pulse times of arrival (\acsp{ToA}) with underlying models allows us to precisely extract the star’s rotational properties, astrometric and, if it has a companion, its orbital characteristics and how these evolve over time. In the case of recycled pulsars, which are typically found in binary systems, five fundamental orbital parameters can be measured with incredible precision~\citep{Lorimer2008}. As outlined below, for a small (but increasing) number of binary pulsars, the timing precision is sufficient to measure additional post-Keplerian parameters allowing mass constraints and potentially future $I$ constraints~\citep{Kramer:2021jcw}. 

For isolated, slowly rotating pulsars that are not in binaries, we cannot measure orbital or post-Keplerian parameters. Instead, radio pulsar timing provides the means to compare \acsp{ToA} with a smooth background model of the \ac{NS}'s regular spin-down behaviour. In particular, frequent timing observations with daily or several-day cadence allow the detection of sudden spin-up glitches (visible as a characteristic cusp-like signal in the timing residuals) in many young pulsars~\citep{espinoza2011}. In addition, regular monitoring of isolated sources also highlights variations in the emission properties, pulse profiles and the stellar spin-down rate~\citep[][see also~\citealt{Oswald2025_MAG}]{Lyne2010, Brook+2016pks, Shaw:2022mxv,Basu:2024qwt, lower_etal_25}, which have been identified as key characteristics of free precession~\citep{jones_andersson_02}. Pulsar glitches and precession provide not only insights into global \ac{NS} properties but also additional information on internal physics on smaller scales.

In the following sections, we discuss recent advances in measuring \ac{NS} masses, moments of inertia, and maximum rotation frequencies as well as detections of pulsar glitches and our understanding of \ac{NS} precession. We will highlight corresponding constraints on sub-atomic physics and touch on several ways that the SKA will provide new insights in each area before providing a detailed overview of how the SKA will contribute to novel dense matter constraints in Section~\ref{sec:SKA_expect}.


\subsection{Pulsar mass measurements}
\label{sec:mass}

\begin{figure*}[t!]
\centering
\includegraphics[width=0.95\linewidth]{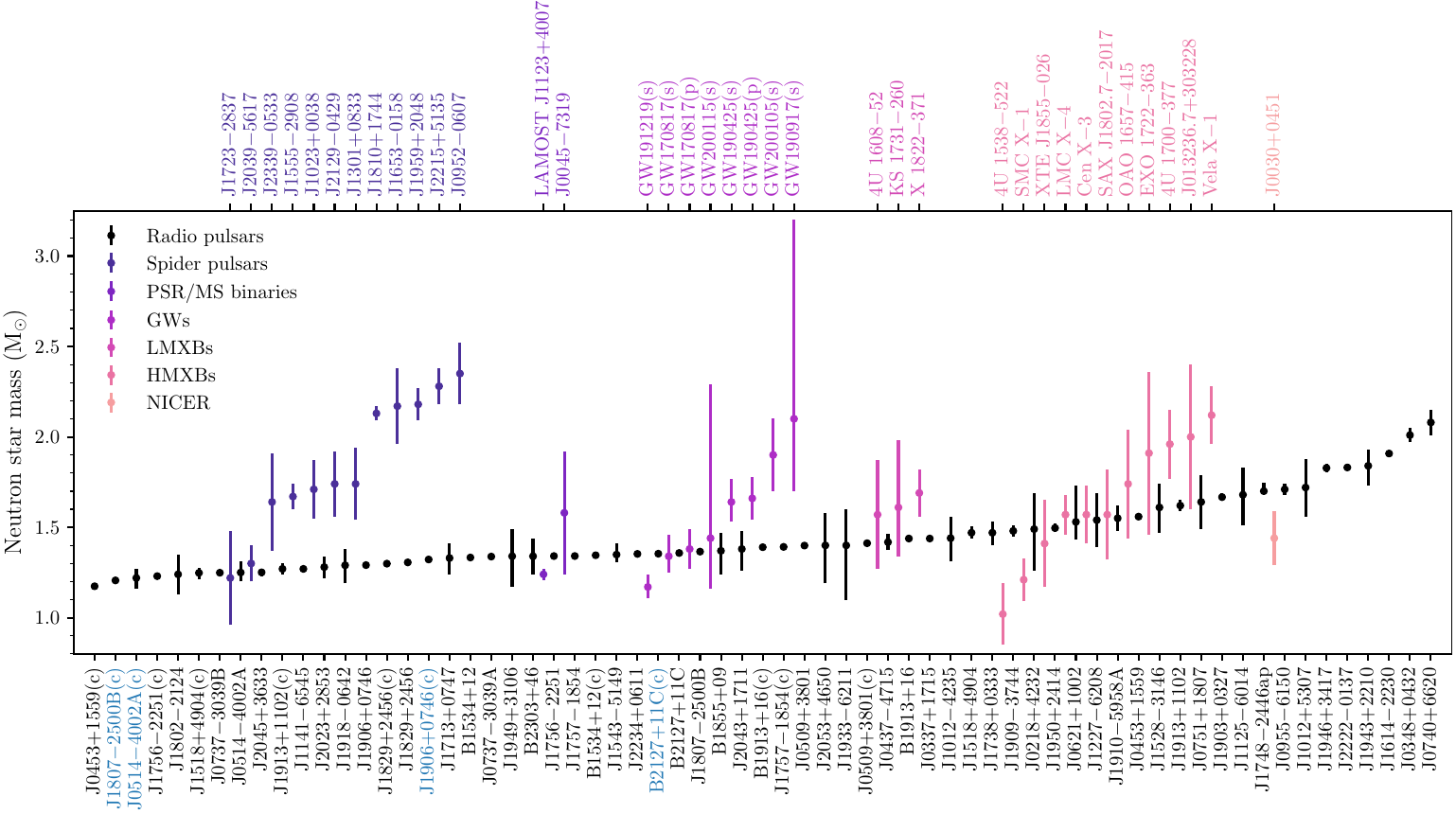}
\caption{The observed \ac{NS} mass spectrum with 68\% confidence intervals sourced from Extended Data Tables 1 and 2 of \cite{You2025} plus updated radio timing measurements (see \url{https://www3.mpifr-bonn.mpg.de/staff/pfreire/NS_masses.html}). The latter are shown in black. Other data points are redback and black widow systems (spider pulsars; dark purple), observations of pulsars with main-sequence companions (PSR/MS binaries; purple), gravitational wave events (GWs; light purple), low-mass X-ray binaries (LMXBs; magenta), high-mass X-ray binaries (HMXBs; pink) and NICER X-ray pulse profile modelling results (light pink). \acp{NS} indexed with a `(c)' indicate the companion to an observed pulsar. Note that those with names in blue could be either a \ac{NS} or a white dwarf based on current constraints. For \acp{NS} detected via \ac{GW} merger events, `(p)' indicates the primary or heavier object and `(s)' the secondary or less massive object.}
\label{fig:NS_masses}
\end{figure*}

To first post-Newtonian order, Einstein's equations for the relativistic effects in the orbital motion and the propagation of radio waves observed in the timing of pulsars in binaries depend only on the masses of the components and the observable Keplerian parameters of the orbit~\citep{Damour1992}. In particular, the rotational stability of millisecond pulsars allows us to measure one or more of the five post-Keplerian timing parameters (i.e., the periastron advance $\dot{\omega}$, the rate of orbital period decay due to \ac{GW} emission $\dot{P}_{\rm b}$, the combined time dilation and gravitational redshift $\gamma$, and the Shapiro delay `range' $r$, and `shape' $s$) to high precision. For this reason, radio timing of pulsars in binaries (and, in one case, of a pulsar in a stellar triple system; \citealt{Ransom2014})---the type of observations also to be undertaken by the SKA---is our primary tool to measure spin frequencies~\citep[e.g.,][]{Basu:2021pyd} and masses of \acp{NS} and white dwarfs very precisely~\citep[e.g.,][]{Guo2021, ColomiBernadich:2024A&A}. Such timing measurements have also enabled precise tests of gravity theories~\citep[see][and~\citealt{VenkatramanKrishnan2025_SKA_Gravity}]{FreireWex2024}. The SKA's future involvement in sensitive VLBI observations will further provide precise parallax measurements essential for calibrating
distance-dependent contributions to timing parameters relevant for mass measurements and tests of gravitational models~\citep{Kramer:2021jcw}.

Over the last 15 years, radio pulsar timing has dramatically broadened the distribution of \ac{NS} masses. 61 corresponding mass measurements with a relative 1$\sigma$ uncertainty of less than 15\% are shown in black in Figure~\ref{fig:NS_masses}. We note that some of these pulsar-timing based measurements have been further enhanced through observations of diffractive scintillation induced by the scattering of radio waves in the interstellar medium (ISM)~\citep{Rickett1990}. The orbital motion of a pulsar can be imprinted as variations in the measured scintillation timescale, which allows us to uniquely identify the sense of the orbital inclination and further refine the pulsar and companion masses, as has been done for PSRs~J1141$-$6545 and J0437$-$4715~\citep{Reardon:2019MNRAS, Reardon:2020ApJ}.

\begin{figure*}[t!]
\centering
\includegraphics[width=0.9\linewidth]{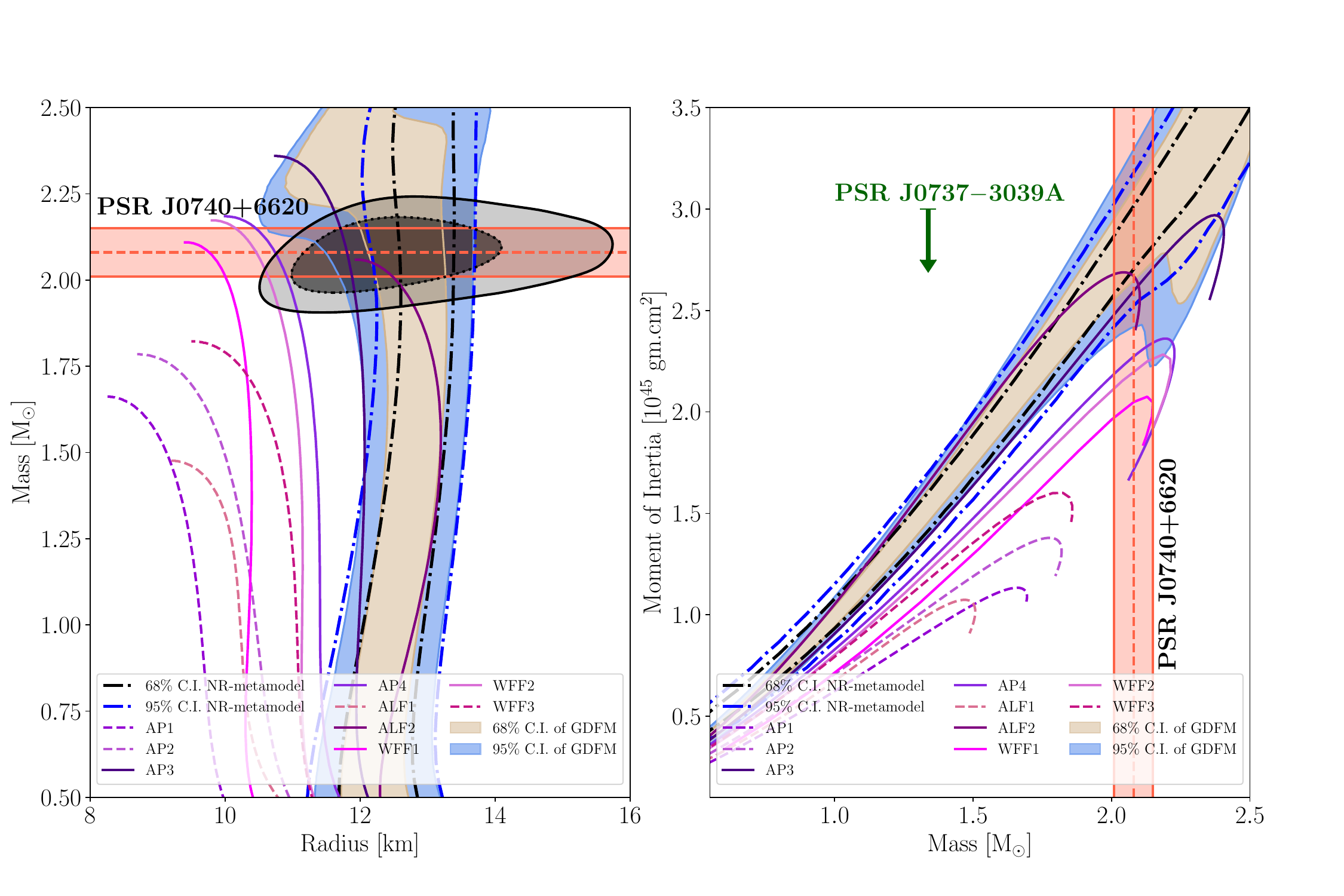}
\caption{\ac{NS} mass-radius relations (left panel) and mass-moment of inertia relations (right panel) for different nuclear \acp{EoS}. The solid lines correspond to the $M$-$R$ and $M$-$I$ sequences for specific \ac{EoS} models described in \citet{LattimerPrakash2001} that satisfy the condition of having a maximum mass exceeding 2.08(7) $M_\odot$, as measured for PSR~J0740$+$6620~\citep{Fonseca:2021wxt} and shown by the orange shaded regions on both plots. The dashed lines represent the $M$-$R$ and $M$-$I$ sequences of \ac{EoS} models that do not meet this maximum mass requirement. The brown and blue shaded regions represent the $M$-$R$ and $M$-$I$ ranges corresponding to the 68\% and 95\% quantiles of the posterior distribution from a Bayesian inference based on a relativistic meta-model representation of the \ac{EoS}~\citep{Char:2023fue}. For comparison, the dash-dotted black and blue contour lines represent the same quantiles obtained from an equivalent analysis with casual and chemically stable instances for a non-relativistic meta-model~\citep{montefusco_2025}. On the left plot, the grey shaded region, bounded by solid black lines, represents the 95\% confidence bounds on the mass and radius of J0740+6620 from NICER measurements, while the inner dashed black line denotes the 68\% confidence interval~\citep{Salmi24a}. The green downward arrow in the left panel shows the 90\% upper limit for $I$ of PSR~J0737$-$3039A using the data from \cite{Kramer:2021jcw}.}
\label{fig:EOS_mass-radius}
\end{figure*}

In general, the radio timing observations summarised in Figure~\ref{fig:NS_masses} now confirm that some \acp{NS} have masses $\gtrsim 2 M_\odot$. Presently, the most massive \ac{NS} determined via pulsar timing is PSR~J0740$+$6620 with a mass of $2.08 \pm 0.07$\,M$_{\odot}$~\citep{Fonseca:2021wxt}. This measurement has had a major impact on the study of ultra-dense matter as any \ac{EoS} that is incapable of sustaining the most massive \ac{NS} observed can be excluded. Consequently, PSR~J0740$+$6620's mass measurement has ruled out the softest \acp{EoS} like those shown by dashed lines in the left panel of Figure~\ref{fig:EOS_mass-radius}. The parameter space for viable \acp{EoS} can be further refined by combining mass estimates with radius or moment of inertia measurements. As the former is inaccessible through radio timing, and the latter difficult to extract (see Section~\ref{sec:MoI} for details), complementary constraints on radii and tidal deformabilities from observations in the X-rays and from \ac{GW} mergers, respectively, are invaluable despite their lower precision. As discussed in more detail in Section~\ref{sec:synergies}, the synergy between X-ray and radio observations is particularly noteworthy in this regard as highlighted by the black contours in the left panel of Figure~\ref{fig:EOS_mass-radius} showing PSR~J0740$+$6620's combined mass-radius measurements with NICER~\citep{Salmi24a}.

In light of future SKA observations, we also highlight that, more recently, the operation of MeerKAT has enabled a large increase in the number and precision of radio timing measurements of binary pulsar masses in the Southern hemisphere~\citep{Serylak2022,Berthereau:2023aod,Shamohammadi2023,Geyer2023,Gautam2024,Bernadich2024,Padmanabh2024,Chisabi:2025ohh}. Currently, the most massive pulsar detected in the Southern part of the sky is PSR~J1614$-$2230 with a mass of $1.97 \pm 0.04$\,M$_{\odot}$~\citep{Demorest2010}. Once the SKA begins taking data, the impact of these measurements will continue to grow as the timing baselines and corresponding precision increase and new pulsars will be discovered~\citep{Keane2025_SKA_Census}.

Additionally, \ac{NS} mass measurements can be obtained from so-called `spider' systems shown in dark blue on the top left of Figure~\ref{fig:NS_masses}. Spiders are binary systems comprising a millisecond pulsar and a low-mass, non-degenerate companion star, typically with orbital periods of less than a day~\citep{Dodge:2024nvl}. Based on the mass of the companion, these spiders are categorised as `black widows' with extremely low mass companions ($M_c < 0.05 M_\odot$) and `redbacks' with higher companion mass ($M_c \gtrsim 0.1 M_\odot$)~\citep{Roberts:2013}. Mass measurements in spiders either rely on the observation of eclipses (primarily in the gamma-rays) or a combination of radio timing with optical spectroscopy. Corresponding mass determinations in the latter scenario are, however, often overestimated~\citep{Clark:2023owb} because they rely on very difficult estimates of the orbital inclinations from multi-wavelength light curves of the irradiated companions, which are heavily dependent on models of the heat distribution on their surfaces~\citep{Voisin2020}.

Despite these uncertainties, Figure~\ref{fig:NS_masses} illustrates that spider pulsar systems appear to host some of the heaviest \acp{NS}~\citep{Linares:2019aua}. These mass estimates hence suggest that significantly more massive \acp{NS} might be detected via radio timing with the SKA in the near future. Such high \ac{NS} masses have also yet to be confirmed via detections of \ac{GW} events (shown in light purple in the top middle of Figure~\ref{fig:NS_masses}). While the issue with \ac{GW} mass measurements is currently one of precision, which will be resolved as more sensitive \ac{GW} detectors come online~\citep{Abac:2025saz}, possible questions regarding the interpretation of massive systems could remain. In particular, several compact objects have now been identified through \ac{GW} merger detections (e.g., the secondary object in GW190814; \citealt{Abbott:2020ApJ}) as well as one via pulsar timing (i.e., the companion of PSR~J0514$-$4002E; \citealt{Barr:2024wwl}) which have masses that reside in the so-called `lower mass-gap' between $\sim 2-3\,M_{\odot}$. In this region, it is unclear whether these sources are \acp{NS} or black holes. Consequently, refined timing and improved mass measurements for known massive \acp{NS} and those that will be identified by the upcoming SKA pulsar surveys discussed in Section~\ref{sec:large_surveys} will provide crucial information on the maximum predicted \ac{NS} mass (the Tolman-Oppenheimer-Volkoff mass; \citealt{Tolman1939, Oppenheimer1939}). This will elucidate the nature of these mass-gap objects and ultimately set tight constraints on the validity of a variety of \ac{EoS} models. Accurate measurements of the maximum mass a \ac{NS} can attain are not only relevant for standard nucleonic \acp{EoS} but are particularly useful to inform the viability of a possible deconfinement phase transition in \ac{NS} interiors or the existence of twin stars~\citep[e.g.,][]{Glendenning:1998ag, Blaschke:2019tbh}.

Besides maximum mass constraints, dense matter properties are also informed by the lightest \acp{NS}. The \ac{NS}'s minimum mass not only affects the underlying core-collapse formation process~\citep{Yasin:2018ckc, Janka2023, Muller2025} and binary evolution~\citep{Tauris2019, You2025} but also the viability of \acp{EoS} models, particularly non-nucleonic ones. Currently, the lightest \ac{NS} candidate observed through radio timing is the companion of PSR~J0453$+$1559 with a mass of $1.174 \pm 0.004\,M_{\odot}$~\citep{Martinez2015}, which is sufficiently high to allow for a wide range of \ac{EoS} families. \ac{NS} minimum masses below $1\, M_{\odot}$ might exist, however, as recently suggested by modelling of the X-ray spectrum of the central compact object in the supernova remnant HESS~J1731$-$347~\citep{Klochkov2015, Doroshenko2022}. While such low minimum masses would have implications for the \ac{NS} composition~\citep{Brodie2023, Sagun2023}, the mass measurement for HESS~J1731$-$347 is based on several assumptions (in particular the source's distance and the underlying atmospheric model) and larger masses can explain the data equally well if the spectral model is adapted~\citep{Alford2023}, highlighting the significant uncertainties of such measurements. Ultimately, future radio timing observations with the SKA will be essential to determining the entire \ac{NS} mass distributions.

Finally, we note that while the mass measurement techniques outlined above (apart from model-dependent spectral X-ray fitting) only allow constraint of \ac{NS} masses that are located in binary systems, pulsar spin-up glitches are potential (albeit also model-dependent) tools to infer masses of isolated, slowly rotating \acp{NS}~\citep[see, e.g.,][and Section~\ref{sec:glitches}]{Ho_2015_science, Pizzochero2017}.


\subsection{Moment of inertia measurements}
\label{sec:MoI}

Above, we illustrated how the TOV equations define a unique relationship between two macroscopic stellar quantities---mass and radius---for a given \ac{EoS}. Astrophysical observations, such as those from PSR~J0740$+$6620~\citep{Fonseca:2021wxt}, thus constrain a region in the mass–radius diagram, thereby placing limits on viable \ac{EoS} models. However, while pulsar timing provides accurate information about spin frequencies and masses, other bulk parameters such as the radius and the moment of inertia are difficult to extract and lower precision measurements of the radius using X-ray data~\citep[e.g.,][]{Riley:2019yda, Miller:2019cac, Salmi24a, Choudhury:2024xbk} are generally used to constrain the \ac{EoS} further. Similarly, the right-hand panel of Figure~\ref{fig:EOS_mass-radius} illustrates the relationship between mass and moment of inertia~\citep{Hartle_1968}. In analogy with the $M$–$R$ plane, one can envisage observational constraints on the $M$-$I$ plane, which would offer independent means of probing the \ac{EoS}~\citep{Lattimer2005}.

A general approach to estimating $I$ involves gamma-ray observations of pulsars. Because a significant portion of the pulsar’s spin-down power, given by $\dot E = 4\pi^2 I \dot P/P^3$, is emitted as gamma rays, measured luminosities can give insights into the \ac{NS} moment of inertia. However, as the gamma-ray luminosity represents only a fraction of the total spin-down energy, this method yields a lower bound on $I$, typically $> 10^{45}\,$g$\,$cm$^2$. Nonetheless, such constraints are weak due to uncertainties in the distance to the pulsar and the beaming geometry, which make precise moment of inertia estimates from gamma-ray data difficult.

It is, however, promising that constraints in the $M$-$I$ plane can (albeit difficult), in principle, be derived solely from high-precision radio timing, one of the key advantages of upcoming SKA observations. In particular, for some systems the timing precision is such that next-to-leading order (NLO) effects (i.e., those beyond standard post-Keplerian effects) become detectable. The foremost example of this is the `double pulsar' PSR~J0737$-$3039A/B. In this system, NLO effects can be detected in the Shapiro delay, aberration, and rate of advance of the orbit's periastron, $\dot{\omega}$~\citep{Kramer:2021jcw}. The latter is the most relevant to us: One of the two measurable NLO contributions to $\dot{\omega}$ is due to relativistic spin-orbit coupling causing the orbital plane to precess about the total angular momentum vector, resulting in $\dot{\omega}_{\rm LT}$. This effect, also known as Lense-Thirring precession, depends on the angular momentum of pulsar A in this system. As the spin of pulsar A is extremely well known---and aligned with the orbital angular momentum~\citep{Ferdman2013}---we can constrain $I$ directly from $\dot{\omega}_{\rm LT}$. Current measurements using $16\,$yr of double pulsar timing data have resulted in an upper limit of $3 \times 10^{45}\, \rm g \, cm^{2}$ at 95\% confidence level on the moment of inertia of PSR~J0737$-$3039A~\citep{Kramer:2021jcw}, which will be dramatically improved upon by SKA-Mid observations~\citep[][see also below]{Hu:2020MNRAS}.

However, improvements in timing precision alone do not address the major roadblock to obtaining accurate $I$ measurements with SKA observations. Systematic uncertainties in the inferred post-Keplerian parameters, namely the orbital period derivative, $\dot{P}_{b}$, from incomplete modelling of acceleration in the Galactic potential can limit the ultimate precision with which the effects of Lense-Thirring precession can be measured. This is presently the major factor limiting moment of inertia measurements with the highly-relativistic double \ac{NS} PSR~J1757$-$1854, for which the effects of Lense-Thirring precession is expected to be even stronger than the double pulsar~\citep{Cameron:2017ody, Cameron:2023pfr}. Better models of the Galactic potential based on ongoing analyses of data produced by large-scale astrometry missions such as Gaia~\citep{Sanderson:2016ApJ} may further reduce or even eliminate these uncertainties over the coming decade, setting the stage for major improvements on moment of inertia constraints in the SKA era.

Moreover, as discussed in detail in Section~\ref{sec:large_surveys}, the SKA will inevitably find more double \ac{NS} systems, and potentially even pulsar black-hole binaries~\citep{Keane2025_SKA_Census, Levin2025_SKA_NSpop}, some of which are likely to be more compact than presently known pulsar binaries. Such discoveries would open up new, independent avenues of measuring the \ac{NS} moment of inertia. 

For completeness, we also highlight that radio timing of pulsars in binaries is not the only way to constrain $I$. As outlined in Section~\ref{sec:glitches}, observing pulsar glitches provides information on fractional moments of inertia of \acp{NS}, i.e., the fractional contributions of the internal superfluid moments of inertia relative to the total stellar moment of inertia~\citep{Graber2018ApJ, Basu:2018fbt, Montoli:2021act, Antonopoulou:2022rpq, Antonelli:2023vpd}. Although these estimates are strongly model dependent, they provide a unique view into the \ac{NS} interior otherwise hidden from view, which are especially powerful when combined with the global constraints outlined above.


\subsection{Maximum spin frequency}
\label{sec:max_spin}

The theoretical maximum spin frequency of a \ac{NS} is equal to the Keplerian frequency, $f_K$, at the stellar surface, i.e., the maximum spin frequency a star can support before mass shedding. To obtain the Keplerian frequency for a given \ac{EoS}, we can perform full \ac{GR} calculations from a general Kerr-like metric. Such efforts have been made by~\citet{Komatsu:1989zz,Friedman1992,Cook:1993qr,Stergioulas:1994ea}.

\begin{figure}[t!]
\centering
\includegraphics[width=0.47\textwidth]{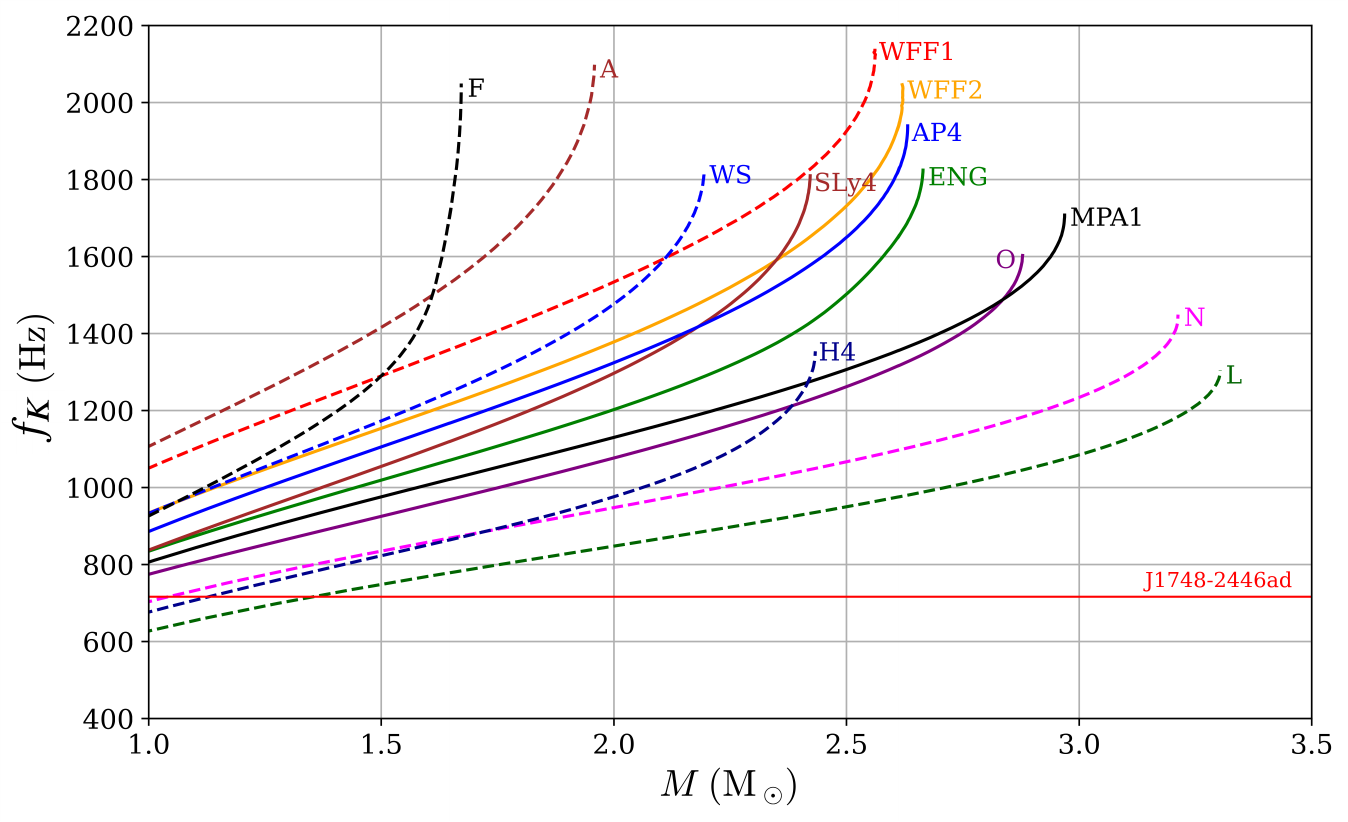}
\caption{Maximum spin frequency for a \ac{NS} as a function of its mass for different \acp{EoS}. The calculations were made with the Rotating Neutron Star (RNS) code (\url{https://github.com/cgca/rns}). \ac{EoS} names are as listed by \cite{LattimerPrakash2001} and in the repository for the RNS code, see also \cite{Stergioulas1996}. The solid curves represent families of maximally rotating NSs from \acp{EoS} that can produce slowly rotating NS configurations with masses consistent with that of PSR~J0740+6620 \citep{Fonseca:2021wxt} and other multi-messenger constraints as discussed in \cite{Dietrich2020}. Dashed lines represent those \acp{EoS} that do not meet the constraints. The red horizontal line shows the spin period of the fastest-spinning pulsar known, PSR~J1748$-$2446ad~\citep{Hessels2006}. For this system no mass has been measured. Higher spin frequencies, especially for systems with well-measured masses, have the potential to constrain the \ac{EoS} in the near future. Calculations and figure by Norbert Wex.}
\label{fig:mass_spin}
\end{figure}

Commonly, the maximum spin frequency of a \ac{NS} is expressed in terms of $M$ and $R$ as follows:
\begin{equation} \label{kepler_freq}
f_K = C \, \left( \frac{M}{M_\odot}  \right)^{\gamma_1} \left( \frac{R}{ 10 \, {\rm km}} \right)^{\gamma_2} .
\end{equation}
The exponents are set to $\gamma_1=0.5$ and $\gamma_2=-1.5$ in the above studies, while different values for the constant $C$ exist in literature. For example, \citet{Haensel:1989mvc} set $C\sim 1225$, and \citet{Friedman1992} use $C\sim 1209$, whereas $C\sim 1045$ in~\citet{Lattimer:2004pg} and $C\sim 1080$ in~\citet{Haensel2009}. As demonstrated in~\cite{Gartlein:2024cbj}, reproducing hybrid stars requires $C$ to no longer be a constant but rather a function that depends on the deconfinement phase transition such that the two limiting cases of hadronic and quark matter are correctly reproduced. Moreover, note that values for $C$ as well as the exponents $\gamma_1$ and $\gamma_2$ are generally chosen in an \ac{EoS} independent way, which is an oversimplification. As different \acp{EoS} predict quite different compactness values, we expect a more compact star (supported by softer \acp{EoS}) to start `mass-shedding’ at a higher frequency. This is highlighted in Figure~\ref{fig:mass_spin}, where we also include several \acp{EoS} that have already been excluded by the large \ac{NS} mass measurements discussed previously for illustration purposes. The figure also demonstrates results from a recent study by~\citet{Basu:2024rio}. Using an agnostic \ac{EoS} parametrisation the authors show that for a large set of \acp{EoS}, stiffer \ac{EoS} realisations (which support larger NS masses) correspond to lower Kepler frequencies. An alternative effort of analysing the maximum frequency of rotating \acp{NS} was performed with the help of a pseudo-Newtonian potential~\citep{Bagchi:2008su}. $f_K$ was found to be lower for both quark and neutron matter \acp{EoS} compared to the limit derived from the \ac{EoS}-independent analytical expression given in Equation~\eqref{kepler_freq}. This reduction in maximum spin frequencies could hint at a general lack of sub-millisecond pulsars.

To date, the observed spin frequencies of millisecond pulsars lie well below the maxima allowed by realistic \acp{EoS} for the full range of \acp{NS} masses. However, it is important to highlight that the largest spin frequency currently measured for a radio pulsar, $716\,$Hz for PSR~J1748$-$2446ad located in the globular cluster Terzan 5~\citep{Hessels2006}, may already exclude several \ac{EoS} models if its mass were known to be small. This statement is supported by the $M$-vs-$f_K$ relation shown in Figure~\ref{fig:mass_spin}. It is evident that the stiffest \ac{EoS} (green dashed curve labelled L) cannot support a \ac{NS} of mass $\sim 1.3 \, M_\odot$, spinning maximally at a frequency of $716\,$Hz. Thus, if the mass of PSR~J1748$-$2446ad was measured to be smaller than this limit, then this specific \ac{EoS} would be ruled out.

Despite being located in a binary system, the prospects of a mass measurement for PSR~J1748$-$2446ad are slim, because it is located in a spider system where relativistic effects are undetectable given the orbital characteristics of the system \citep{Hessels2006}. Complementary estimates at other wavelengths are also unlikely given the current non-detection of the system at optical or near-infrared wavelengths. However, as discussed below in detail, it is possible that future SKA pulsar searches~\citep{Keane2025_SKA_Census} will find some equally fast or faster-spinning pulsars in systems where mass measurements via radio timing are possible. In such cases, the \ac{EoS} could be significantly constrained even for spin periods below $1000\,$Hz. Thus, as Figure~\ref{fig:mass_spin} shows, the combination of a fast spin with a \ac{NS} mass measurement is a more powerful probe of the \ac{EoS} than a fast spin period alone.

While measuring the \ac{NS} spin frequency does not require it to be in a binary system, the companion is crucial for determining its mass. This naturally leads to an intriguing question: Is there a particular type of binary system where \acp{NS} are both exceptionally massive and rapidly spinning? Interestingly, spider pulsar systems, which are known to host some of the heaviest \acp{NS}~\citep{Linares:2019aua} as discussed in Section~\ref{sec:mass} and illustrated in Figure~\ref{fig:NS_masses}, may provide the answer. Combining the criteria for spider systems from Section~\ref{sec:mass} with the condition of a spin period $ < 16\,$ms (a rotational cut-off identifying millisecond pulsars based on the work by~\citealt{Halder:2023rfu}), we filter millisecond pulsars from the ATNF Pulsar Catalogue~\citep[][\href{https://www.atnf.csiro.au/research/pulsar/psrcat/}{https://www.atnf.csiro.au/research/pulsar/psrcat/}]{manchester:2005atnf}. Dividing them further into two groups, spider pulsars and non-spider millisecond pulsars, we find that the median spin frequency of the former group is approximately 1.3 times higher than that of the latter population, with a positively skewed distribution. This indicates that spider pulsars have higher probability of hosting both massive and fast spinning pulsars. Although mass measurements in these systems rely on uncertain optical light curve modelling (see Section~\ref{sec:mass} for details), their discovery---followed by precise radio timing---may provide the orbital parameters needed to model the light curves and determine the \ac{NS} masses~\citep{Dodge:2024nvl}. The SKAO, with its exceptionally sensitive telescopes, will thus play a crucial role in classifying and precisely timing numerous spider systems.

In practice, most \acp{NS} are likely spinning well below their Keplerian frequency, as various mechanisms can act to halt the spin-up process caused by accretion in LMXBs, which are considered the progenitors of millisecond radio pulsars. Specifically, the spin distribution of \acp{NS} in LMXBs appears bimodal, with a more widely distributed `slow' population, and a narrow `fast' population, with a cut-off at around 700 Hz~\citep{Patruno17}. This is very different from the spin frequency distribution of millisecond radio pulsars, which shows no hints of bimodality. Moreover, spin frequencies of millisecond pulsars cut off well below any predicted Keplerian frequencies calculated based on very general grounds assuming only a low density \ac{EoS} and causality for the denser \ac{NS} regions~\citep{HaskellZdunik18}. Additional torques are therefore required to explain the observed distribution, and \ac{GW} torques have been shown to be good candidates~\citep{Gittins19}. Furthermore, observations of transitional sources (see \citealt{Papitto2022} for a recent review) that exhibit both X-ray emission during the accretion phase, and radio emission in quiescence, may shed light on \ac{GW} emission mechanisms~\citep{HaskellPatruno17}.

In fact, the lack of radio pulsars in the lower left quadrant of the $P$-$\dot{P}$ plane (rapidly rotating with low spin-down rates) has been shown to be well modelled in terms of \ac{GW} torques, due to a small residual ellipticity possibly held in place by a buried superconducting magnetic field~\citep{Woan18}. A more detailed understanding of the spin distribution of the millisecond pulsar population---a key science goal for the SKA~\citep{Keane2025_SKA_Census}---will, hence, allow us to understand not only the evolution of magnetic fields and the electromagnetic emission of \acp{NS} in the SKA era, but also the evolution of their quadrupoles and \ac{GW} emission, informing \ac{GW} searches with current and future ground based interferometers~\citep{Haskell15}.


\subsection{Pulsar glitches} 
\label{sec:glitches}

\begin{figure*}[t!]
\centering
\includegraphics[width=0.95\linewidth]{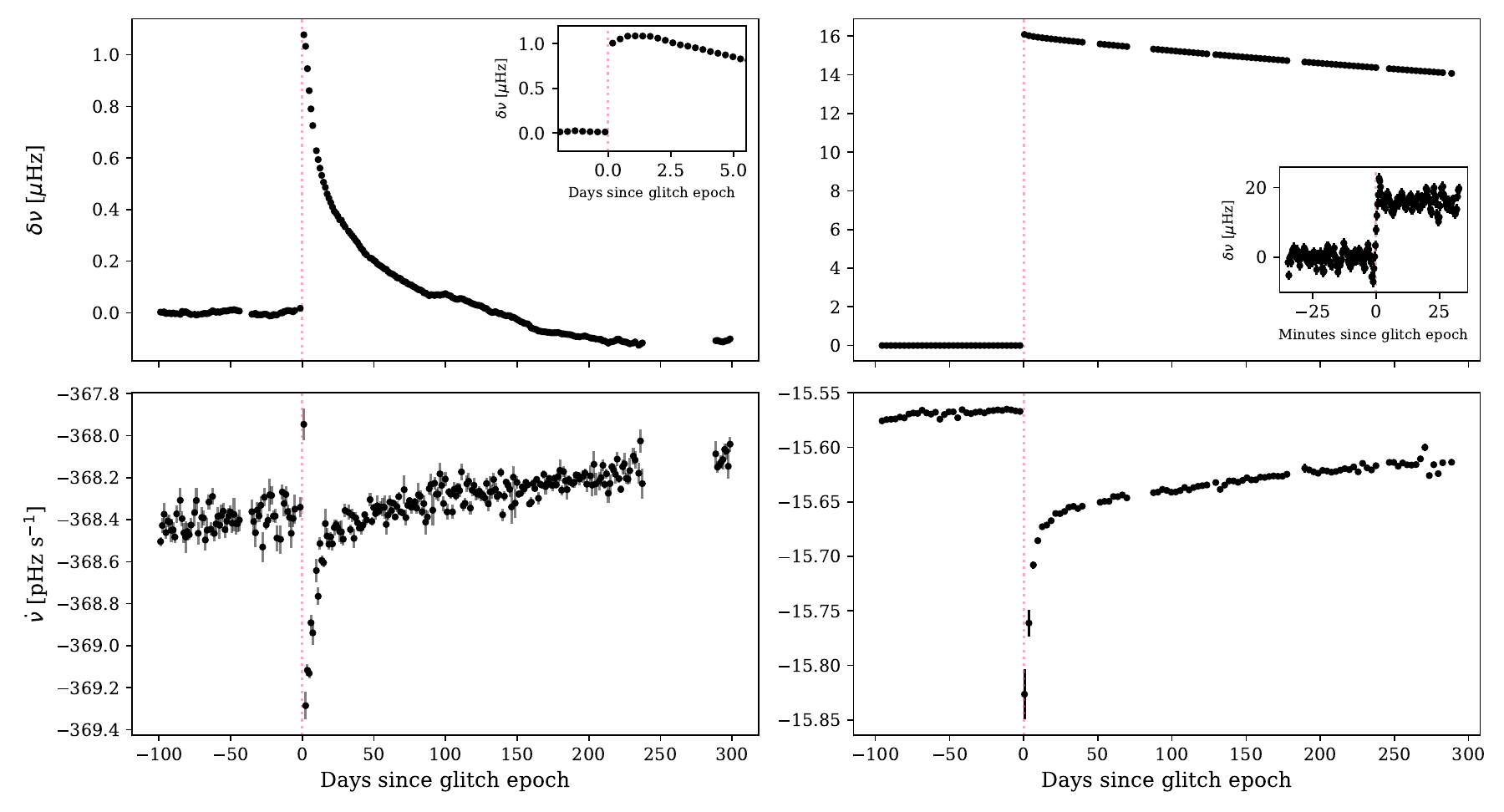}
\caption{Evolution of the spin-frequency (upper row) and the spin-down rate (lower row) of PSR~B0531$+$21 (the Crab pulsar, left column) and PSR~B0833$-$45 (the Vela pulsar, right column) close in time to their respective large glitches in 2019 and 2016. The values were computed using a striding boxcar method (e.g., \citealt{Shaw:2018MNRAS}) in which a fit for both parameters was applied to consecutively overlapping groups of times-of-arrival. The insets show the more densely sampled frequency evolution around the glitch epoch (vertical dotted line).}
\label{fig:crabvelaf0f1}
\end{figure*}

Glitches are sudden increases in a pulsar's rotational spin frequency, which are associated with the rapid angular momentum transfer from the superfluid interior to the observed component of a \ac{NS}~\citep{Antonopoulou:2022rpq,Zhou:2022}. As discussed in Section~\ref{sec:unknowns_SFphysics}, this transfer occurs when many previously pinned vortices unpin and migrate outwards, spinning down the superfluid while accelerating the \ac{NS} crust~\citep{haskell2015IJMPD,Antonopoulou:2022rpq} leading to an observed step $\Delta\nu>0$. In most events, the spin-down rate $|\dot{\nu}|$ also increases abruptly ($\Delta\dot{\nu}<0$), as internal components decouple, and the relaxation of the crust-interior system is reflected in the observed rotation for a duration referred to as `post-glitch recovery'. These sudden spin-ups provide access to dense-matter physics, as glitch characteristics can reveal the \ac{NS} structure (e.g., the amplitudes for the $\nu$ and $\dot{\nu}$ changes depend, among others, on the relative moment of inertia between the various components), internal transport properties and microphysical processes (which, for example, dictate the relaxation timescales). See Section~\ref{sec:unknowns_SFphysics} for further details.

Glitches present a range of amplitudes with $10^{-10}\lesssim\Delta\nu\lesssim10^{-5}\,$Hz\footnote{See the \ac{JBO} Glitch Catalogue~\citep[][\href{https://www.jb.man.ac.uk/pulsar/glitches.html}{https://www.jb.man.ac.uk/pulsar/glitches.html}]{Basu:2021pyd}.} and timescales that vary from seconds for the spin-up, up to years for the subsequent post-glitch recovery. This diverse phenomenology is exemplified by two glitches of the Crab and Vela pulsars in Figure~\ref{fig:crabvelaf0f1}. The frequency residuals (top panels) and the evolution of the spindown rate $\dot{\nu}$ (bottom panels) illustrate their differences in glitch amplitudes, spin-up timescales (zoomed-in inset panels), and recoveries. Whilst Crab's post-glitch spindown rate quasi-exponentially relaxes to its pre-glitch projected value about a year later, in Vela changes persist beyond an initial exponential decay and up to the next glitch, leaving long-lasting effects on its rotation. Despite the often striking differences seen even in the glitch behaviour of a single pulsar, qualitative and quantitative trends for the glitch activity across the pulsar population have emerged as the number of recorded glitches grows~\citep{espinoza2011,fuentes2017,lower:2021mnras,Basu:2021pyd}. To disentangle the many uncertain physical parameters governing glitches it is important to have high-quality observations, both to discover more glitches and to obtain the unique detailed description of individual events. The SKA, with a combination of high-cadence timing of selected glitching pulsars and large-scale population monitoring, is ideally suited for this task and could transform glitches into precise probes of fundamental \ac{NS} properties.

Most glitch rises are unresolved by current observations. In three Vela glitches, the spin-up occurred within tens of seconds (e.g., Figure~\ref{fig:crabvelaf0f1}, top right) and was immediately followed by a fast (minute-long) internal response likely originating from the star's core~\citep{Palfreyman2018,Ashton:2019Nat,Graber2018ApJ,pizzochero2020}. However, for six of the largest glitches of the Crab pulsar, an initial unresolved spin-up was observed to be followed by an extended phase lasting many hours (e.g., Figure~\ref{fig:crabvelaf0f1}, top left)~\citep{lyne:1992nat,wong:2001apj,ge:2020apj,Basu:2019iam, Shaw:2018MNRAS,Shaw:2021MNRAS}. This rare feature of a slow rise is also seen once in each of the radio-quiet magnetars 1E~2259$+$568~\citep{woods:2004apj} and SGR~J1935$+$2154~\citep{ge:2022ar}. The detection of glitch rises in the Crab and Vela pulsars is, to a large extent, possible due to the observing resources afforded to these NSs~\citep{lyne:2015mnras,Dodson2007}. Such observations of the spin-up and subsequent shortest relaxation features (seen in Vela) inform us of the dynamics of superfluid vortices inside \acp{NS}~\citep{AntonelliHaskell2020,sourie:2020mnras} and the role of the star's core in glitches~\citep{Graber2018ApJ,haskell:2018mnras}, while also providing critical constraints on crustal entrainment and the moment of inertia of superfluid components~\citep{andersson2012PhRvL,chamel2013PhRvL,Montoli:2018fqz,Montoli:2020A&A}. 

\begin{figure}[t!]
\centering
\includegraphics[width=0.477\textwidth]{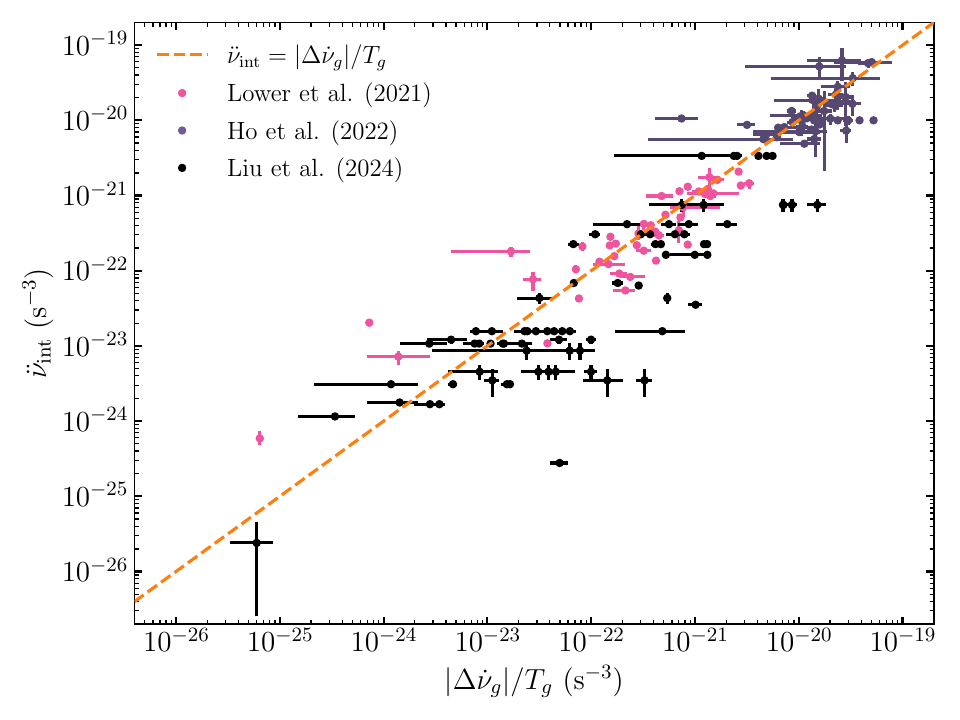}
\caption{Observed relationship between the second spin-frequency derivative measured between glitches ($\ddot{\nu}_{\rm int}$) and the glitch-induced step-change in spin-down rate normalised by the wait time to the next glitch ($|\dot{\nu}_{g}|/T_{g}$). Points in pink are those presented in \citet{lower:2021mnras}, purple ones are from \citet{Ho:2022kga} for PSR~J0537$-$6910, and in black are the values from \citet{Liu2024} (note, the $\ddot{\nu}$ values from this work are the average across multiple glitches). The dashed orange line indicates the one-to-one relationship which would suggest a complete recovery of $\dot{\nu}$ at the time of the next glitch.}
\label{fig:glt_f2}
\end{figure}

Regular observations will facilitate studies of long-term relaxations, while also increasing the number of detected glitches. The former can shed light on the dynamics of weakly-coupled superfluid components, point to possible internal conditions governing the observed variability in glitch phenomenology~\citep{celora:2020mnras,haskell20}, and advance our understanding of the influence of glitches on the overall rotational evolution of young pulsars. The latter, i.e., the discovery of more glitches in typical glitching \acp{NS} as well as millisecond pulsars, is pivotal for statistical studies. The distributions of glitch magnitudes and inter-glitch waiting times vary from pulsar to pulsar~\citep{melatos:2008apj, fuentes2017}. Some---like the Crab---show signs of a random process, yet others---like Vela---present a characteristic size and waiting time which can be used to infer the properties of the angular momentum reservoir and constrain microphysical parameters such as vortex pinning forces, entrainment, and the superfluid gap model~\citep{andersson2012PhRvL,Ho_2015_science}. Uniquely, PSR~J0537$-$6910's next glitch can be predicted by the size of its previous one, indicative of a threshold-dominated process as the glitch trigger~\citep{Antonopoulou:2017hwa,Melatos2018ApJ}. The distributions and correlations for individual pulsars, as well as for glitches collectively (e.g., the relation of glitch activity with pulsar inferred ages), form an essential input for understanding  the conditions under which glitches occur~\citep{Antonopoulou:2022rpq,Antonelli:2023vpd}. For example, Figure~\ref{fig:glt_f2} presents a  recently established correlation for a subgroup of glitching pulsars, which relates their glitch-induced changes in spin-down rate to waiting times~\citep{lower:2021mnras,Liu2024} and offers new test ground for glitch models. Currently, such studies are limited by the small sample size, an area that SKA observations can considerably improve (see Section \ref{sec:glitchobs}). 

To fully leverage SKA's capabilities and maximise the scientific gains, a coordinated observational strategy is essential (detailed in Section~\ref{sec:SKA_expect}), with flexibility to change which sources take priority for high-cadence observations, e.g., when a glitch happens or is expected for a particular pulsar.


\subsection{Free precession}
\label{sec:precession}

Free precession offers another point of connection between SKA pulsar observations and the \ac{NS} structure. If modelled as a simple biaxial rigid body, a misalignment between the star's (fixed) angular momentum vector and its symmetry axis results in a periodic modulation of the rate and latitude at which the magnetic axis sweeps around the angular momentum vector (see, e.g.,~\citealt{Jones2001,gao_etal_23}). If the moment of inertia tensor is diagonal and has components $(I_x, I_x, I_z)$, then, for small angle free precession of a nearly spherical star of spin period $P$,  a long-period modulation $P_{\rm mod}$ is induced in all aspects of the pulsar emission, i.e., in the pulse timing, beam shape, and polarisation, with 
\begin{equation}
    \frac{P}{P_{\rm mod}} \approx \frac{|I_z - I_x|}{I_z} \equiv \epsilon .
\end{equation}
The precession modulates the rate at which the magnetic axis swings past the observer's line of sight, i.e., modulates the observed spin frequency. It also modulates the angle between the star's (fixed) angular momentum vector and the pulsar beam, resulting in variations in the angle through which the observer's line of sign cuts the pulsar beam.  This will produce variations in the beam shape and the observed sweep of polarisation with rotational phase.  Also, providing (as is likely) the spin-down torque is sensitive to this latitudinal angle, there will be variation in the spin-down torque, amplifying the variations in spin frequency~\citep{cordes_93,Jones2001}. 

The modelling of \ac{NS} crusts suggests that the ellipticity is limited to values $\epsilon \lesssim 10^{-6}$~\citep{Jones2001, Johnson2013, Gittins21a, Gittins21b}, so that a typical $P \sim 1$\, second pulsar would display modulations on the timescale of months. However, the connection between observation and theory is complicated by the realisation that if a pinned superfluid component exists, as required to explain pulsar glitches (see Sections~\ref{sec:unknowns_SFphysics} and~\ref{sec:glitches}), then the modulation period is drastically reduced~\citep{Shaham1977}:
\begin{equation}
    \frac{P}{P_{\rm mod}} \approx \frac{I_{\rm SF}}{I_z} ,
\end{equation}
where $I_{\rm SF}$ is the moment of inertia of the pinned superfluid component, believed to be of the order a few percent of the total stellar moment of inertia. It follows that the observable modulations might appear on timescales as short as tens of seconds for a typical $P\sim 1$\,s pulsar. Furthermore, in this scenario, the precession may itself be damped rather rapidly, due to the dissipative interaction of the neutron superfluid vortices with the star's magnetic field~\citep{link_06}. Clearly, there is a wide range of theoretical possibilities, demanding the analysis of SKA data with both high cadence and long duration.

A number of pulsars have already been seen to display quasi-periodic oscillations in their spin-down rates, with free precession sometimes being advanced as the cause, the cleanest example being PSR~B1828$-$11~\citep{Stairs2000, Ashton2016, Ashton2017}. However, subsequent analysis has revealed that in some cases at least, these timing variations are accompanied by sharp changes in emission profile, something not expected for free precession~\citep{Lyne2010, Stairs2019}. It has been argued that this may reflect the pulsar magnetosphere~\citep{Oswald2025_MAG} being finely balanced between two states, with the changing geometry of the free precession providing a statistical bias as to which state is preferred at any given precessional phase~\citep{Jones2012}.  However, the origin of the timing variations themselves remains unclear.

Further progress on understanding the origin of these quasi-periodic oscillations will require analysis of contemporaneous spin-down and pulse profile data, ideally for a large set of pulsars. This is particularly important given the recent results of~\citet{lower_etal_25}, who found that such correlated changes may be more common than previously thought. Carrying out this analysis for as large a set of pulsars as possible is crucial, as the statistical correlation (or lack thereof) between the pulsars' spins and modulation periods potentially contains important information on the mechanism at work in producing the periodicity~\citep{Jones2012}.

Furthermore, there are potential connections between free precession and the glitches described in Section~\ref{sec:glitches}. First, as noted above, the storage of pinned vorticity in a neutron superfluid may produce extremely short modulation timescales~\citep{Jones:2016oyh}.  Second, it is possible that non-axisymmetries in pulsar glitches may themselves `kick' the star into free precession~\citep{jones_andersson_02}, with a crust-cracking event suddenly changing the moment of inertia tensor, instantaneously producing a misalignment between the (fixed) angular momentum vector and a principal axis.  This opens up the need to record timing data in the time interval \emph{immediately following} a glitch, something which the SKAO should be well placed to achieve.


\section{Potential impact of dark matter and modified gravity models}
\label{sec:DM_modgrav}

Given that little is known about one quarter of the total energy of the Universe, the dark sector might affect pulsar observations. While this field of research is relatively new, studies show that the impact of Beyond the Standard Model physics and gravity beyond \ac{GR} show a degeneracy with \ac{EoS} uncertainties, hindering us from probing the baryonic matter \ac{EoS}~\citep{Yazadjiev:2014cza,Giangrandi:2022wht}. 

This is relevant because due to their extreme compactness, \acp{NS} can accumulate a sizeable amount of \ac{DM} particles from the surrounding galactic medium. This is especially important for the observations of pulsars close to the Galactic centre (see also~\citealt{Abbate2025_SKA_GalCen}), where the \ac{DM} density is significantly higher than far away from the centre, a feature referred to as the \ac{DM} spike~\citep{Ullio:2001fb}. 

\begin{figure*}[t!]
\centering
\includegraphics[height=0.45\linewidth]{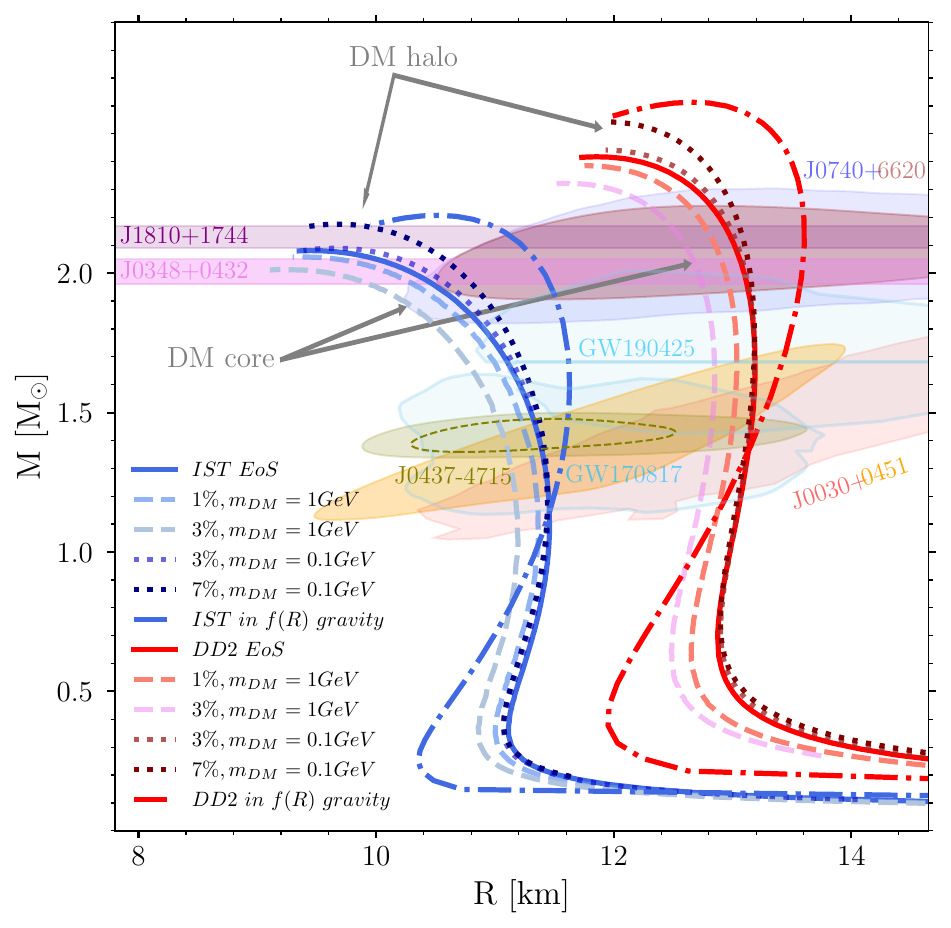}
\includegraphics[height=0.45\linewidth]{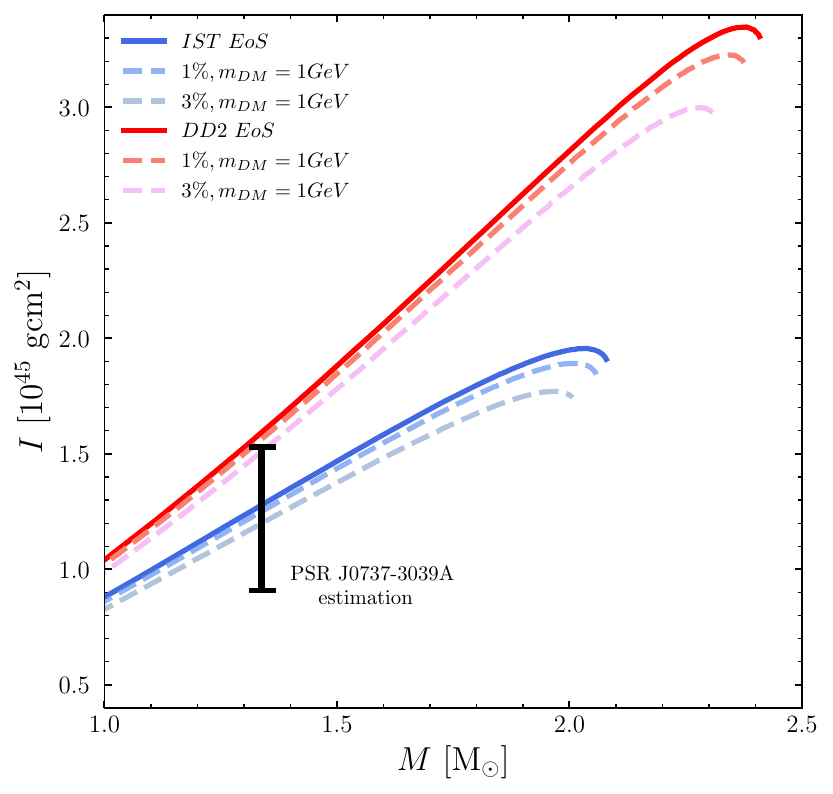}
\caption{Total gravitational mass as a function of the visible baryonic radius (left panel) and the moment of inertia as a function of the total gravitational mass (right panel) for \ac{DM}-admixed \ac{NS} is shown for several \ac{DM} fractions, and \ac{DM} particle masses, $m_{\rm DM}$. The results were obtained for asymmetric fermionic \ac{DM} that interacts with the visible sector only through gravity (for more details see~\citealt{Ivanytskyi:2019wxd}). To address the baryonic matter \ac{EoS} uncertainties, we consider the soft IST \ac{EoS}~\citep{Sagun:2020qvc} (solid blue curve) and stiff DD2~\citep{Typel:1999yq} (solid red curve) as two extreme limits~\citep{Cipriani:2025tga}. The dashed and dotted curves depict the core and halo configurations, respectively, considering \ac{GR}. The $M(R)$ dash-dotted blue and red curves were obtained for the $f(R)= R+\alpha R^{2}$ gravity for $\alpha$=100~\citep{Yazadjiev:2014cza, Yazadjiev:2015zia}. The 1$\sigma$ confidence interval constraints from GW170817, GW190425, the NICER measurements of PSR~J0030$+$0451~\citep{Miller:2019cac, Vinciguerra:2023qxq}, PSR~J0740$+$6620~\citep{Dittmann24,Salmi24b}, and PSR~J0437$-$4715~\citep{Choudhury:2024xbk}, as well as mass measurements of heavy radio pulsars (PSR~J1810$+$1744, PSR~J0348$+$0432) are also plotted. The estimated $I$ of PSR~J0737$-$3039A (without accounting for \ac{DM} or modified gravity) with 90\% confidence interval is depicted in black in the right panel~\citep{Landry:2018jyg}.
}
\label{fig:DM}
\end{figure*}

If the interaction between the visible and dark sectors other than gravity is very weak, the components do not exist in equilibrium. In this case, the \ac{DM}-admixed NSs are defined by two pressures and two energy densities, one for each of the components. Additionally, the amount of \ac{DM} in each star may vary depending on the \ac{DM} density in the surrounding medium, which results in a two-dimensional plane, i.e., \ac{DM} fraction and particle mass or interaction strength ~\citep{Hippert:2022snq}. 

The impact on the \ac{NS} properties, such as gravitational mass, radius, moment of inertia, matter distribution, etc., depends on the \ac{DM} candidate, its particle mass, and self-interaction strength~\citep{Bramante:2023djs,Cipriani:2025tga}. Very weakly interacting heavy \ac{DM} particles tend to form a dense core inside the baryonic \ac{NS} leading to a smaller radius for a given gravitational mass of the star, which is related to the additional gravitational pull of \ac{DM}~\citep{Leung:2011zz,Ivanytskyi:2019wxd,RafieiKarkevandi:2021hcc,Giangrandi:2022wht,Barbat:2024yvi}. This effect mimics the softening of the \ac{EoS} caused by the appearance of new heavy degrees of freedom including deconfined quarks~\citep{Biesdorf:2024dor}. On the other hand, light \ac{DM} can form an extended halo around the baryonic \ac{NS} that increases the star's total gravitational mass and resembles the stiffening of the \ac{EoS}~\citep{Nelson:2018xtr,Karkevandi:2021ygv}. A similar effect is obtained for axions and other ultra-light particle clouds surrounding a \ac{NS}~\citep{Noordhuis:2023wid}. The corresponding $M(R)$ relations for different \ac{DM} configurations and the impact of \ac{DM} on $I$ are shown in Figure~\ref{fig:DM}.

In this \ac{DM} analysis, \ac{GR} was assumed. However, it has been shown that a modification of gravity itself can also impact the \ac{NS} properties and, consequently, inferred quantities such as mass and radius~\citep{Shao:2019gjj}. As illustrated in the left panel of Figure~\ref{fig:DM}, the $M(R)$ relations obtained for \ac{GR} and $f(R)=R+\alpha R^2$ gravity~\citep{Starobinsky:2007hu} exhibit a high degree of degeneracy making it difficult to disentangle the effects. Therefore, measuring only the \ac{NS} radius at a given mass would not be sufficient to distinguish between the effects of \ac{DM} or modifications of gravity and the dense matter properties.

Although \acp{NS} provide a compelling testing ground for gravity, nuclear physics, and physics Beyond the Standard Model, the possible degeneracy between the effects of \ac{DM} and/or gravity beyond \ac{GR} and dense matter properties could lead to misleading conclusions when analysing SKA data in isolation. Therefore, the joint efforts of the SKA and other facilities to obtain multi-messenger observations of \acp{NS}, along with advances in experimental and theoretical subatomic physics are pivotal for breaking these possible degeneracies and shedding light on the \ac{NS} internal composition.


\section{Expectations in the SKAO era}
\label{sec:SKA_expect}

The SKAO will comprise two telescopes, SKA-Low and SKA-Mid, which will be the largest and most sensitive radio telescopes at cm wavelengths in the Southern Hemisphere. We refer to 
\href{https://www.skao.int/en/science-users/599/scientific-timeline}{https://www.skao.int/en/science-users/599/scientific-timeline} for details on both telescopes. Their discoveries will push the boundaries of fundamental science. However, in this paper, we confine our discussion to the SKA's impact on advancing our understanding of ultra-dense matter around and beyond nuclear saturation density.


\subsection{Advances due to SKAO's improved sensitivity}

The sensitivity of the SKA telescopes will progressively increase as more antennas are deployed, significantly enhancing their scientific capabilities over the next decade. For comparison, the SKA-Mid arrays will be around three and four times more sensitive than MeerKAT in the AA* and AA4 configurations, respectively. A similar progression is expected for SKA-Low, which will eventually surpass the sensitivity of the low-frequency array LOFAR by nearly an order of magnitude once fully deployed.

This substantial boost in sensitivity for the different array configurations will directly translate into reduced uncertainties on pulsar \acsp{ToA}. As a result, precision measurements of masses in both existing and newly discovered binary systems will become possible at an unprecedented level. Given that MeerKAT has achieved mass measurements with a precision of $\sim4 - 20$\% from Shapiro delay measurements alone~\citep{Corongiu:2023gft, Shamohammadi2023, Berthereau:2023aod, Geyer2023, Gautam2024, Jang:2024A&A, ColomiBernadich:2024A&A, Grunthal:2024A&A}, the SKA-Mid array will bring drastically improved precision on mass measurements, enabling the tightest constraints on the maximum \ac{NS} mass to date. The high sensitivity and broad frequency coverage of SKA-Mid will also facilitate complementary constraints on the pulsar orbital parameters through scintillation-based measurements. 

\begin{figure}[t!]
\centering
\includegraphics[width=0.48\textwidth]{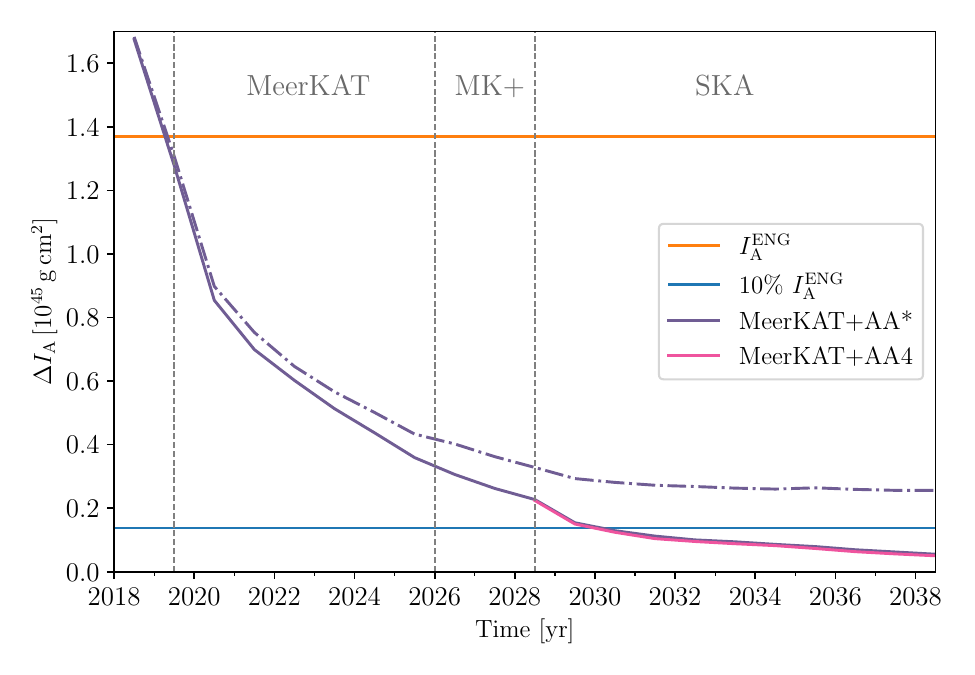}
\caption{Simulated uncertainty for a moment of inertia measurement for pulsar A in the double pulsar system J0737$-$3039 as a function of observing time with MeerKAT, MeerKAT+ (\href{https://www.meerkatplus.tel/}{https://www.meerkatplus.tel/}) and SKA. All simulations are based on the ENG \ac{EoS}~\citep{LattimerPrakash2001} and $3\,$hr monthly observations with the full array configurations. The horizontal orange line marks the theoretical value for pulsar A's moment of inertia, $I_{\rm A}$, while the blue horizontal line shows an uncertainty of 10\%. The purple dash-dotted line shows the evolution of the measurement uncertainty of $I_\mathrm{A}$ based on recent Galactic measurements~\citep{Gravity2021, Guo2021}. It decreases with time and eventually levels off from 2030 onwards. In contrast, the solid lines (purple: AA*, pink: AA4) show predictions assuming no errors in the Galactic parameters justified given that these are expected to improve in the future. Improvements in the timing parallax and proper motion are also included in the analysis.}
\label{fig:MoI_uncertainty}
\end{figure}

The increase in sensitivity will similarly set unprecedented constraints on the \ac{NS} moment of inertia. The most promising existing candidate is the double pulsar PSR~J0737$-$3039A/B (see right panel of Figure~\ref{fig:EOS_mass-radius}). To estimate corresponding uncertainties with time, we can simulate the expected precision of the three relevant post-Keplerian parameters (the advance of periastron $\dot{\omega}$, the Shapiro parameter $s$, and the rate of change of orbital period $\dot{P}_\mathrm{b}$) with MeerKAT and the SKA (see~\citealt{VenkatramanKrishnan2025_SKA_Gravity} for details on these simulations). Subsequently removing the extrinsic kinematic contributions and performing the $\dot{P}_\mathrm{b}$--$\dot{\omega}$--$s$ test using the intrinsic contributions for the former two quantities as outlined in~\citet{Hu:2020MNRAS} leads to uncertainties in the moment of inertia for pulsar A of $10-23$\% by 2030 and $4-20$\% by 2038 (after 10 years of timing with the SKA) at 68\% confidence, respectively. These ranges are determined by our uncertain knowledge of the underlying Galactic potential as discussed in Section~\ref{sec:MoI} and highlighted in Figure~\ref{fig:MoI_uncertainty}.

The key for enabling new mass and moment of inertia constraints is the availability of observing resources. Measurements as those outlined above require us to time corresponding pulsar systems for at least several hours each month with SKA-Mid to build up suitable observing baselines. We note that while SKA-Low is expected to detect a wealth of new pulsars across the entire sky, sources of special interest will need to be followed up with SKA-Mid to mitigate the deleterious effects of pulse scatter broadening and variable dispersion due to propagation through the ISM at low observing frequencies. 

\subsection{The role of large surveys}
\label{sec:large_surveys}

Pulsar science, and consequently constraints on ultra-dense matter, in the SKA era will not only advance through increases in sensitivity and future observations of currently known systems but also new wide-field and targeted programs. The SKA's exceptional sensitivity across a broad range of frequencies will transform pulsar surveys by enabling the detection of extremely faint, unknown sources across a variety of environments. While low-frequency observations are generally favoured due to the steep emission spectra of pulsars~\citep{Posselt:2022vrk}, which makes them intrinsically brighter at low frequencies, the ISM introduces significant scattering and dispersion, particularly towards the Galactic centre~\citep{Rickett1977}. In this region, SKA-Mid’s higher-frequency capabilities will be crucial for overcoming ISM effects and accessing the dense inner Galaxy, where a large pulsar population is anticipated~\citep{Gonthier2018, Mishra-Sharma2022, Abbate2025_SKA_GalCen}. Consequently, SKA-Mid is expected to perform a Galactic plane survey, with SKA-Low covering the remaining sky~\citep{Keane2025_SKA_Census}. While difficult to predict given the uncertainties of the underlying population synthesis models and specific survey strategies, we can expect these surveys to detect on the order of $10,000$ slow pulsars as well as $\sim800$ millisecond pulsars and $\sim 110$ double \ac{NS} systems in the AA* configuration; AA4 will lead to an increase by roughly a factor 1.2~\citep{Keane2025_SKA_Census}. The SKA is further expected to increase the known population of young pulsars with large spin-down power, i.e., prime candidates for pulsar glitches. Of the currently known $\sim 200$ glitching \acp{NS}, around 140 have characteristic ages below $10^6\,$yr. This subsample is responsible for around $600$ of the $\sim 700$ glitches observed to date~\citep[see][and \href{http://www.jb.man.ac.uk/pulsar/glitches.html}{http://www.jb.man.ac.uk/pulsar/glitches.html} for details]{Basu:2021pyd}. Based on the evolutionary pulsar population synthesis simulations outlined in~\citet{Keane2025_SKA_Census}, we expect around $600$ \acp{NS} with such characteristic ages for AA*, while this number will increase to $\sim 650$ \acp{NS} for AA4, promising a wealth of new glitch observations.

Increasing the known \ac{NS} population will naturally increase the number of systems that can act as unique laboratories for the types of nuclear physics constraints outlined previously. Of the $\sim3,800$ known radio pulsars---a tiny fraction of all Galactic \acp{NS} beamed towards the Earth~\citep{Graber:2023jgz}---only around 10\% are in binaries~\citep[ATNF Pulsar Catalogue, v.2.6,][\href{https://www.atnf.csiro.au/research/pulsar/psrcat/}{https://www.atnf.csiro.au/research/pulsar/psrcat/}]{manchester:2005atnf}. Of these, a few dozen have properties that allow mass constraints (see Figure~\ref{fig:NS_masses}) with only a handful being sufficiently massive to provide relevant constraints on the \ac{EoS}~\citep{Antoniadis:2013pzd, Fonseca:2021wxt}. An increase in sample size will potentially lead to the discovery of additional systems with masses above $2 M_{\odot}$, providing new \ac{EoS} constraints. Targeted follow-up searches of unidentified Fermi gamma-ray sources~\citep{Camilo:2015caa} and globular cluster systems~\citep{Bagchi2025_SKA_GlobClust} in the Galaxy, most susceptible to millisecond pulsars, are particularly relevant in this context. Additionally, mapping out the overall \ac{NS} mass distribution (and reliably determining the minimum \ac{NS} mass) is essential not only for constraining exotic \acp{EoS}~\citep{Sagun2023} and our understanding of \ac{NS} formation through \ac{EoS}-dependent core-collapse supernovae~\citep{Yasin:2018ckc, Janka2023}, but also for compact binary evolution~\citep{You2025} and \ac{GW} signals from binary \ac{NS} mergers~\citep{Harry:2018hke, LIGOScientific:2018cki}.

Similarly, we might be fortunate enough to discover a double \ac{NS} system similar to the double pulsar but with an orbital period of less than one hour. In such a case, the accuracy of a moment of inertia measurement is expected to reach 10\% after six years of timing observations with the SKA, and improve to about 1\% after a decade~\citep{Hu:2020MNRAS}. Moment of inertia constraints might also be possible in a pulsar-black hole system, a class of binaries that has remained elusive to date but whose detection is one of the key science goals for the SKAO~\citep{VenkatramanKrishnan2025_SKA_Gravity}. Given that gravitational effects in these binaries are stronger than for double \ac{NS} systems~\citep{Liu:2014uka}, moment of inertia measurements will likely require shorter observing baselines compared to the double pulsar to reach similar uncertainties~\citep{bagchi2013}. The discovery of new pulsars in large-scale surveys might also reveal more information about the pulsar spin distribution and processes that affect the \ac{NS} rotation~\citep{Levin2025_SKA_NSpop}. In particular, out of the $\sim 3,800$ known radio pulsars around 640 are millisecond pulsars with periods below $30\,$ms, of which $\sim 560$ rotate faster than $10\,$ms~\citep[ATNF Pulsar Catalogue, v.2.6,][\href{https://www.atnf.csiro.au/research/pulsar/psrcat/}{https://www.atnf.csiro.au/research/pulsar/psrcat/}]{manchester:2005atnf}. While this does not yet provide sufficient information to rule out families of \acp{EoS} (see Figure~\ref{fig:mass_spin}), a detection of a sub-millisecond pulsar combined with a mass measurement would set new constraints on the ultra-dense matter \ac{EoS} (see Section~\ref{sec:max_spin} for details).

Predicting the expected number of these new system is challenging, but every single such discovery with the SKA has the potential to transform our knowledge of astrophysical constraints of dense matter.


\subsection{Achievable ToA precision with the SKAO -- A case study}
\label{case_study}

The precision on a \acs{ToA} measurement can be approximated by \citep{Bailes:2018azh} 
\begin{equation} \label{sigamtoa}
    \sigma_{\rm ToA} \approx \frac{W}{2 {\rm (S/N)}},
\end{equation}
where S/N is the signal-to-noise ratio of the integrated (folded) pulse profile, and $W$ is the pulse width. This S/N is proportional to the number of elements used for beamforming. Computing the S/N for SKA-Low is straightforward as every element is identical, whereas for SKA-Mid, a subarray can be assembled by combining different proportions of MeerKAT and SKA-Mid dishes. In such a scenario, S/N can be expressed as
\begin{equation} \label{radiometer}
    {\rm S/N} = S_{\rm mean} \sqrt{\frac{P-W}{W}} \frac{(M G_M + N G_S)^2 \sqrt{N_p \Delta f T_{\rm obs}}}{(M G_M T_M + N G_S T_S)}.
\end{equation}
Here, $S_{\rm mean}$ is the mean radio flux density of the pulsar at a given frequency $f$, $P$ is the spin period of a pulsar, $N_{\rm p} = 2$ is the number of summed polarisations, $\Delta f$ is the bandwidth, $T_{\rm obs}$ is the total integration time of an observation, $M$ is the number of MeerKAT dishes each with antenna temperature gain $G_M$ and system temperature $T_{M}$, and $N$ is the number of SKA-Mid dishes each with antenna temperature gain $G_S$ and system temperature $T_{\rm S}$. The antenna temperature gain of an array element is defined as the ratio of the effective area to twice the Boltzmann constant, $A_{\rm eff}/2k_B$. Equation~\eqref{radiometer} reduces to the standard formula~\citep{lorimerkramerhandbook} when beamforming is achieved through an array in which all elements are of the same type. For all calculations that follow, we adopt values of $G_M = 0.042\,$K/Jy, $T_M = 20$ K~\citep{Bailes:2018azh}, $G_S = 0.058\,$K/Jy and $T_S = 13.5$ K~\citep{Pellegrini+2020} and assume that SKA-Mid observations, which include MeerKAT dishes, are carried out at L-Band\footnote{`L-Band' refers to the range of frequencies around $1420\,$MHz. In the case of the SKA, this is covered by Band 2, which ranges from $950-1760\,$MHz.}. 

In order to estimate the achievable \acs{ToA} precision at each array assembly, we adopt a range of fiducial pulsar parameters. We select a pulse profile width of $2.5\,$ms based on a histogram of pulse widths from all pulsars in the ATNF Pulsar Catalogue~\citep[][\href{https://www.atnf.csiro.au/research/pulsar/psrcat/}{https://www.atnf.csiro.au/research/pulsar/psrcat/}]{manchester:2005atnf}, with the chosen value corresponding to the most populated bin. We further choose representative values of $300\,$ms for the pulse period (corresponding to a rotation frequency of $\nu = 3.3$ Hz) and $-1.8 \times 10^{-12}\,$Hz s\textsuperscript{-1} for the frequency derivative $\dot{\nu}$. We also adopt $S_{\rm mean, 250} = 1.1\,$mJy at $250\,$MHz and $S_{\rm mean, 1400} = 0.046\,$mJy at $1400\,$MHz, approximately corresponding to the flux densities of the faintest known glitching pulsar in the Jodrell Bank Observatory (\ac{JBO}) Glitch Catalogue~\citep[PSR~B1911$+$11;][\href{https://www.jb.man.ac.uk/pulsar/glitches.html}{https://www.jb.man.ac.uk/pulsar/glitches.html}]{Basu:2021pyd}. 

\begin{figure*}[t!]
\centering
\includegraphics[width=0.85\linewidth]{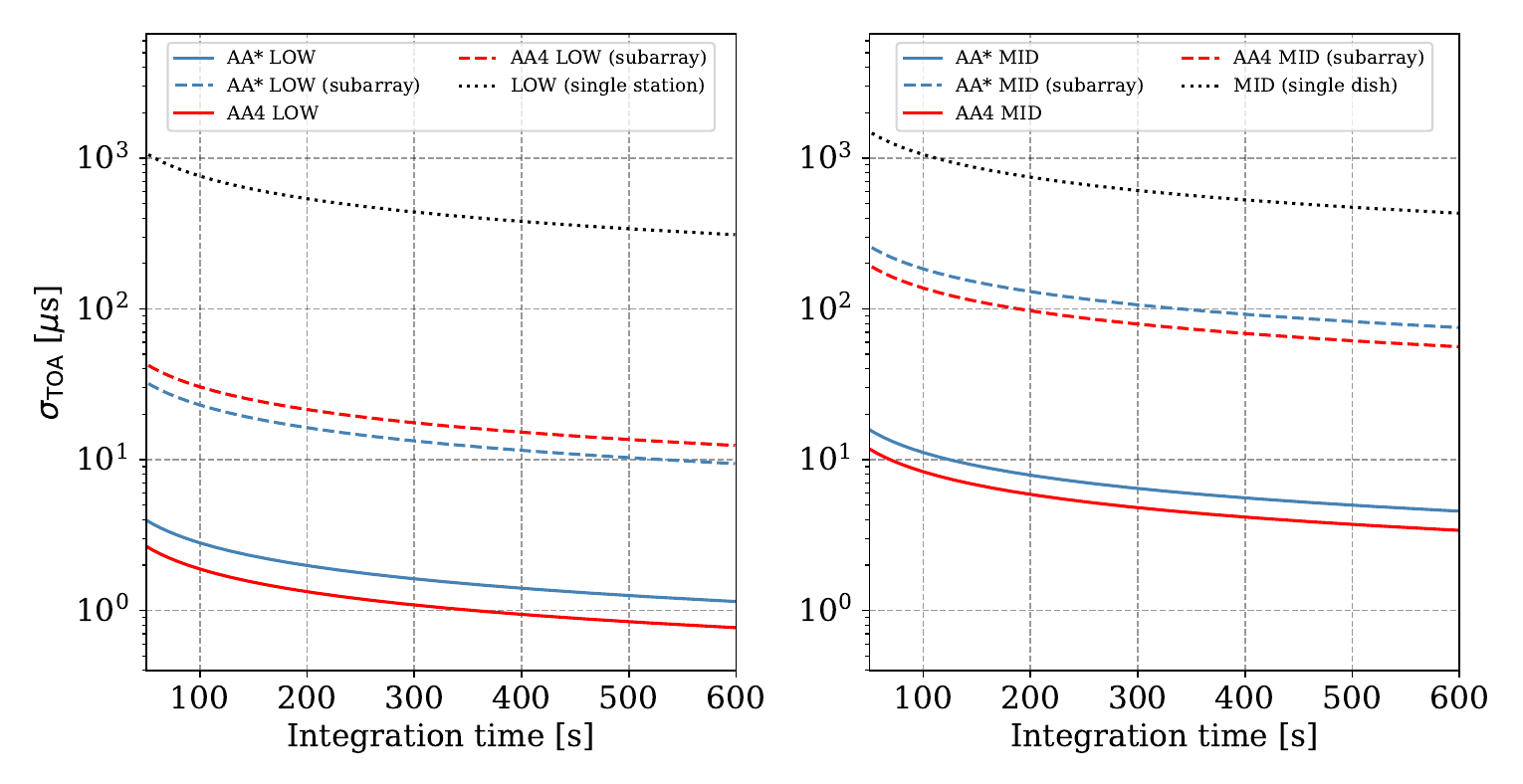}
\caption{\acs{ToA} precision achievable with integration times of up to $600\,$s for a young, faint pulsar with typical properties (see text), shown for SKA-Low (left) and SKA-Mid (right). Solid lines indicate the precision attainable, when observing with all array elements within a radius of 10 km of the core, of AA* (blue) and AA4 (red).  Dashed lines represent the use of a subarray comprising $1/n$\textsuperscript{th} of the elements within 10 km of the core, where $n$ is the number of Pulsar Timing Subsystem (PST) beams
each array assembly can process concurrently (For SKA-Mid, this is 16 beams for both AA* and AA4. For SKA-Low, only 8 beams will be processed at AA*, 16 for AA4). The black dotted line shows the \acs{ToA} precision using a single array element.}
\label{fig:toa_tint}
\end{figure*}

Figure~\ref{fig:toa_tint} shows the expected \acs{ToA} uncertainties for various SKA-Mid and SKA-Low configurations. For SKA-Low, we assume the pulsar to be observed near the zenith. For the canonical pulsar described above, observing with all 271 SKA-Low elements within 10km of the core during AA* yields a timing precision of approximately 2$\mu$s after about 4 minutes of integration. With the 404 stations within 10km of the core in AA4, close to 1 $\mu$s precision can be reached in the same integration time. However, these estimates likely represent lower bounds. In practice, the achievable precision—particularly at lower frequencies—may be degraded by interstellar scatter broadening, which distorts the pulse profile and reduces the effective S/N, especially for distant or highly scattered sources. Additionally, at short-integration times, \acs{ToA} precision will also be limited by pulse-to-pulse jitter. For SKA-Mid, high-precision timing can be achieved through longer integration times, which may be the preferred approach for precise mass and moment-of-inertia measurements, as ISM effects are reduced at higher observing frequencies. Additionally, to maximise timing precision, observations with SKA-Low should favour pulsars with a comparatively low dispersion measure, compared with SKA-Mid.

Although the full array gives impressive sensitivity in terms of \acs{ToA} errors, a compromise on sensitivity is required to observe multiple sources simultaneously. Both SKA-Low and Mid will have the capability to form up to 16 concurrent (subarray) beams for pulsar timing in AA4. For SKA-Mid, 16 beams will also be possible for AA*, but SKA-Low will process up to 8 beams during AA*. This will allow a range of simultaneous observing possibilities as well as the allocation of smaller subarrays, and shorter integration times for brighter pulsars, to facilitate the efficient use of telescope time (e.g., \citealt{song:2021MNRAS}), whilst reducing the effects of pulse jitter. The dashed lines in Figure~\ref{fig:toa_tint} represent the case where the array is split into equal parts according to the maximum number of beams each array assembly can process. In this case, for SKA-Low, AA* has more array elements per beam than AA4 (because the number of beams is relatively  limited), resulting in improved \acs{ToA} precision for SKA-Low during AA*. However, the flexibility of the SKAO means that a subarray can comprise any number of elements from a single dish/station up as many elements as required within a 20 km baseline.

In Figure~\ref{fig:toa_tint}, we also present the case where a single array element is used (dotted lines). In this case, after a 4 minute integration of the pulsar described above, we can achieve $\sigma_{\rm ToA} = 0.5\,$ms for Low and $0.7\,$ms for Mid. However, given that the S/N scales with flux density, and in this case we chose a particularly faint pulsar ($\sim1\,$mJy at 250 MHz), a factor of $\sim$100 improvement in $\sigma_{\rm ToA}$ could be achieved when using single stations to target brighter ($\sim100\,$mJy) pulsars for a similar integration time. 


\subsection{Observing pulsar glitches with the SKAO}
\label{sec:glitchobs}

To probe the full glitch parameter space, an observing program should be devised in such a way that it resolves rotational variations on timescales from seconds to years for a large sample of pulsars. Moreover, as glitches can be seen as transient events, which may require alterations to observing methodology, a flexible approach to glitch detection and characterisation with the SKA is crucial. To optimally use the sensitivity of the SKA to detect small glitches, in both real-time and retrospective searches (e.g., \citealt{melatos:2020ApJ}; \citealt{singha:2021MNRAS}), a high cadence is necessary to distinguish them from timing noise. 

The nominal cadence from routine timing programs is insufficient to capture some medium-term glitch recoveries on timescales below tens of days~\citep{Liu2024}. Resolving these recoveries will require either including the pulsar in a dedicated higher-cadence timing program~\citep{Basu:2019iam, lower:2021mnras, Basu:2021pyd, Liu2024} or employing a glitch detection approach such that cadences can dynamically be temporarily adjusted in order to resolve the recovery phase. 

High cadence observations can be made possible with both SKA-Mid and SKA-Low through flexible approach to scheduling. For instance, when a new glitch is detected during the routine timing program(s), a (subarray) beam can be temporarily allocated to the pulsar to enable improved cadence. In addition, observing cadences can be further enhanced by exploiting commensal observing capabilities, allowing access to many pulsars within the wide field-of-view, even if the main target of the observation is not a pulsar.  Such observations will be valuable for probing the prevalence of glitches and similar rotational features, as well as for discovering rare events such as
antiglitches (abrupt decreases in spin frequency primarily observed in magnetars \citep{Archibald2013}, atypical glitches, and glitches from unusual sources, such as millisecond or very old pulsars. Furthermore, it may be possible to alert other observatories of the occurrence of a glitch\footnote{Similarly, glitches detected by other observatories could be followed up using the SKAO telescopes, where resources allow.} for rapid follow-up if the SKA's flexibility is limited at the time of the glitch event. The attainment of improved timing precision, close in time to a glitch epoch, will also be of interest to \ac{GW} detection facilities~\citep[e.g.,][]{Moragues2023}. 

\begin{figure*}[t!]
\centering
\includegraphics[width=0.85\linewidth]{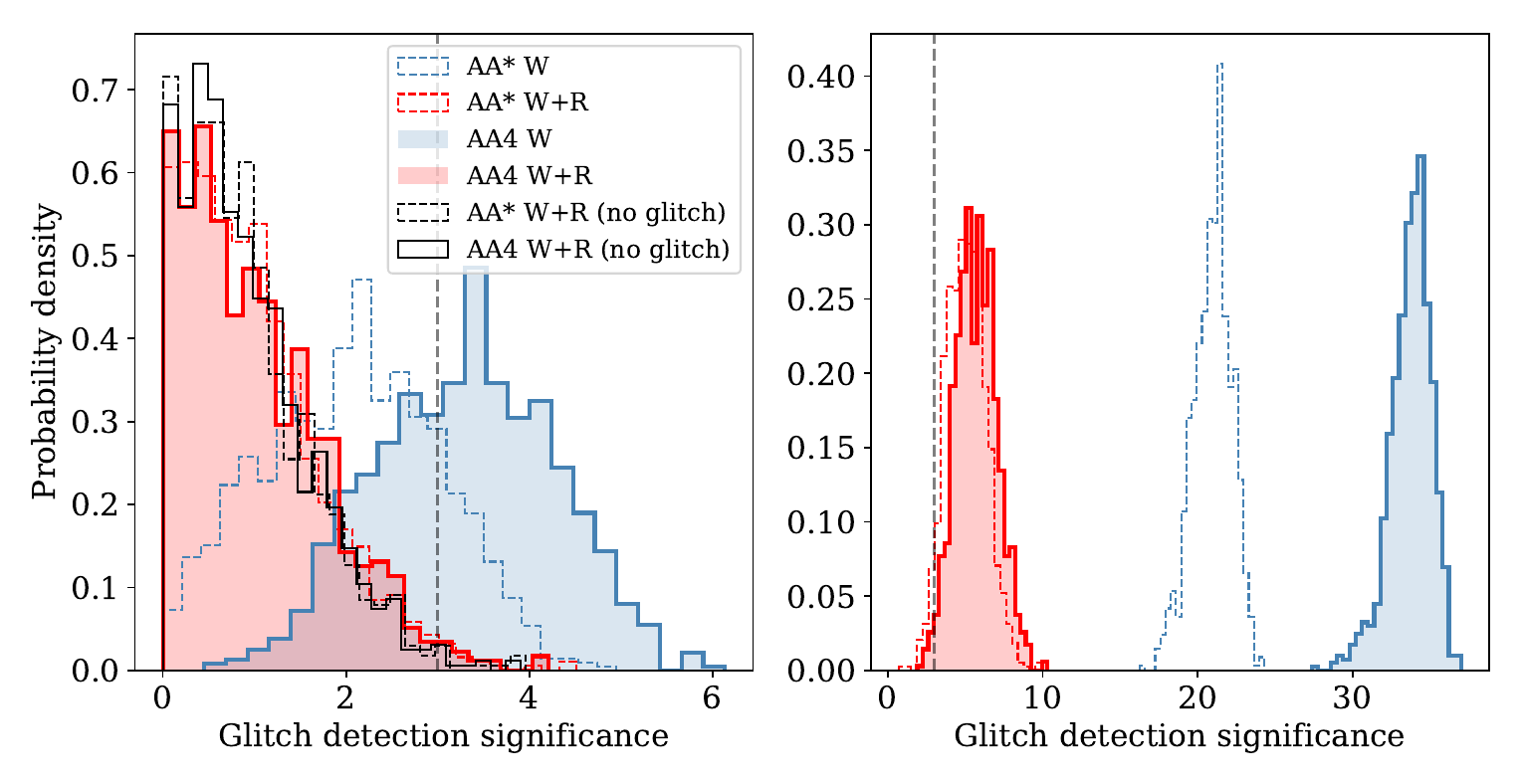}
\caption{Detection significance of simulated pulsar glitches based on the first post-glitch \acs{ToA}, as determined by a real-time glitch identification algorithm (see text). The data assume long-term timing using all available SKA-Low stations for AA* (dashed histograms) and AA4 (filled solid histograms). A nominal observing cadence of one observation every two weeks is used. Each panel shows the distribution of detection significances for simulated glitches with amplitudes of $\Delta \nu = 1.8 \times 10^{-11}\,$Hz (left) and $\Delta \nu = 1.8 \times 10^{-10}\,$Hz (right). Blue curves correspond to \acsp{ToA} affected only by white noise (W), while red curves include additional red noise (W+R) (see text). The black solid (AA4) and dashed (AA*) histograms in the left panel represent detection significances in the case where no glitches were injected into the red noise. The vertical dashed lines denote $3\sigma$ significance. Each of the 1000 simulations per histrogram uses a different noise realisation. Distributions are normalised to unit area. The legend applies to both panels.}
\label{fig:glitch_sig}
\end{figure*}

Resolving the very short-term glitch effects requires a combination of high cadence and sufficiently long integration time to increase the likelihood of observing during and shortly after the event. The sensitivity and large field-of-view of the SKA telescopes are ideal for catching glitches on the rise, revealing
the true prevalence of delayed spin-ups, and enabling studies of the short-term recovery process. Such an approach would also benefit from robust online glitch detection, triggering real-time alerts and diverting any available resources to the pulsar of interest as soon as possible. For instance, during long integration time observations of one of a selected group of bright, young pulsars which are part of a dedicated high-cadence monitoring programme, \acsp{ToA} can be calculated dynamically whenever a predefined S/N threshold is reached, resulting in multiple closely-spaced \acsp{ToA} from the same observation. In the event of a detected glitch, the observing schedule can be automatically re-prioritised to focus resources on the glitched pulsar, ensuring high-cadence coverage to capture the critical, rapid post-glitch evolution. The long integration times required could be achieved through the use of any idle SKA-Low station beams, in combination with commensal observations. This approach would maximise observing efficiency while maintaining the flexibility to respond dynamically to newly detected glitches. Targeted searches for possible magnetospheric emission changes associated with glitches---as recently seen in Vela~\citep{Palfreyman2018}---or for thermal X-ray signatures indicative of heat deposition, as expected if some glitches involve a pulsar crustquake~\citep{Bransgrove2020} will also be made possible by this approach. 

To examine the significance with which the SKA-Low can detect a glitch in near real-time\footnote{In this context, `real-time' refers to the immediate evaluation of new \acs{ToA} as soon as an observation is made to determine whether a glitch has occurred since the previous observation. This contrasts with `retrospective' glitch searches, which analyse longer spans of historical data to identify glitches after the spin-up has taken place.}, we simulate normally distributed timing residuals, into which a glitch is injected, using the pulsar timing package developed in \cite{antonopoulou:2011mnras}. This corresponds to the ideal case in which there is no red noise and the timing residuals are dominated by \acs{ToA} uncertainty. We use nominal observing cadence of approximately one \acs{ToA} every two weeks. We then evaluate whether the glitch can be identified using only the first post-glitch \acs{ToA}. Using an integration time $T_{\rm obs} = 4$ minutes, we achieve $\sigma_{\rm TOA} = 1.8 \mu$s for AA* and $\sigma_{\rm TOA} = 1.1 \mu$s for AA4. To inject the glitch into the timing residuals, we introduce a step change in both $\nu$ and $\dot{\nu}$ near the midpoint of the dataset. The primary glitch amplitude is set to $\Delta \nu = 1.8 \times 10^{-11}$ Hz, corresponding to the smallest absolute value of $\Delta \nu$ recorded in the \ac{JBO} Glitch Catalogue~\citep[][\href{https://www.jb.man.ac.uk/pulsar/glitches.html}{https://www.jb.man.ac.uk/pulsar/glitches.html}]{Basu:2021pyd}.\footnote{This glitch occurred around MJD 57120 in PSR~B0410$+$69.} We then assess whether the first post-glitch timing residual is significantly inconsistent with the expected value, assuming no glitch has occurred, for this amplitude and for a second case with a glitch ten times larger. To evaluate the impact of red noise on glitch detectability, we repeat the simulations of timing residuals using noise drawn from a power-law spectrum~\citep{lentati:2014mnras}, the power spectral density of which is given by,
\begin{equation}
	P(f) = \frac{A_{\rm red}^2}{12 \pi^2} \left( \frac{f_{\rm t}}{1 \mathrm{yr}^{-1}}\right)^{-\gamma}, 
    \label{timingnoise}
\end{equation}
where $f_{\rm t}$ is the Fourier frequency\footnote{The inverse of the timescale of the injected red noise, expressed in cycles per year. The upper and lower bounds of $f_{\rm t}$ are determined by the observing cadence and the timing baseline of the dataset respectively. }. We adopt the typical values of amplitude $A_{\rm red} = 1 \times 10^{-11}$ yr\textsuperscript{$-$3/2} and spectral index $\gamma = -4.9$ for all red noise realisations. For each glitch amplitude and array assembly configuration, we generate 1000 independent simulations of the timing residuals, each with a unique realisation of the white noise, or combined white and red noise. 

To determine the presence or absence of a glitch, and the significance with which it can be detected, we employ the `near real-time' glitch detection approach currently implemented in the \ac{JBO} pulsar timing programme.  This is carried out using Gaussian Process Regression which models a segment of the pre-glitch residuals using a squared-exponential covariance function with an additive white noise kernel. The pre-glitch model (formed using the 100 \acsp{ToA} preceding the glitch epoch) is extrapolated to the time of the first post-glitch observation (the test epoch). We then measure the significance, using a $z$-score statistic, of the value of the residual at the test epoch, compared to this pre-glitch extrapolation. The results of these simulations are shown in Figure~\ref{fig:glitch_sig}. 

The left panel shows results for the smallest glitch (i.e., the smallest $\Delta \nu$) in the \ac{JBO} Glitch Catalogue~\citep[][\href{https://www.jb.man.ac.uk/pulsar/glitches.html}{https://www.jb.man.ac.uk/pulsar/glitches.html}]{Basu:2021pyd}. In the absence of red noise (blue distributions), AA4 (solid) provides a clear advantage over AA* (dashed), with the mean detection significance exceeding $3\sigma$ (indicated by the vertical dashed line). As expected, the presence of red noise (red distributions) substantially reduces the significance of the detection. Notably, for this particular parametrisation of red noise, AA4 offers little improvement over AA*, a trend also visible in the right panel, which depicts a glitch ten times larger. For the smaller glitch, a small number of detections exceed the $3\sigma$ significance level. To assess whether these are likely to be false positives, we ran additional simulations with no glitch injected but with red noise present. The resulting distributions (black histograms: dashed for AA*, solid for AA4) show a similar number of $> 3\sigma$ detections, suggesting that some of the apparent detections in the glitch-injected case may arise spuriously from noise. It is worth noting that the pulsar from which the glitch amplitudes in these simulations were derived (PSR~B0410$+$69) does not exhibit significant red noise in the \ac{JBO} dataset. As such, the white noise simulations provide a more realistic representation of this specific case. When we apply the same detection algorithm to the \ac{JBO} data for PSR~B0410$+$69, the first post-glitch \acs{ToA} yields a detection significance of only $1.5\sigma$, falling short of a confident real-time identification. Our simulations thus demonstrate that, in scenarios with minimal red noise, AA4 provides a clear advantage over both AA* and \ac{JBO} in detecting glitches of this small amplitude, on the timescale of the observing cadence. This is particularly crucial for identifying glitches in millisecond pulsars (e.g., \citealt{cognard:2004apj}; \citealt{mckee:2016mnras}), which tend not to exhibit substantial levels of timing noise. 

\begin{figure}[t!]
\centering
\includegraphics[width=0.47\textwidth]{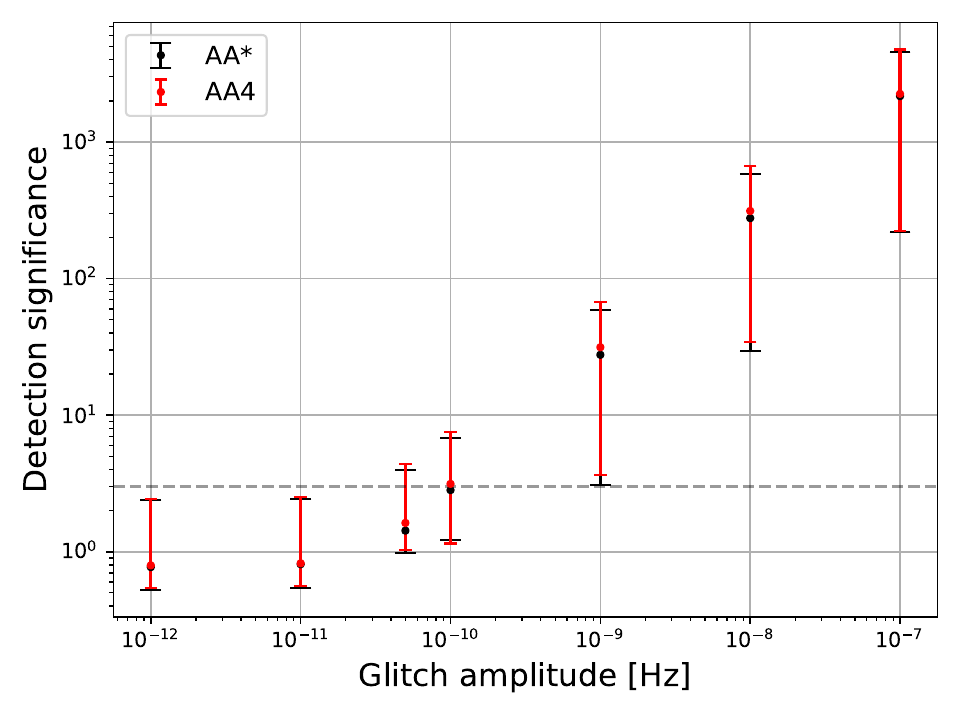}
\caption{Detection significance for a range of simulated glitch amplitudes in which red noise (according to the $A_{\rm red}$ and $\gamma$ values used above) is present in the data. Each point represents the median detection significance for 1000 realisations of the red noise for a given glitch amplitude. Error bars indicate the 68\% confidence interval, spanning the 16th to 84th percentiles of the distribution about the median. The horizontal dashed line denotes the $3\sigma$ level.}
\label{fig:size_sig}
\end{figure}

Figure~\ref{fig:size_sig} extends Figure~\ref{fig:glitch_sig}, showing the glitch detection significance that results from further simulations, all of which include red noise, over a wider range of glitch amplitudes. All other parameters for these simulations were the same as those underpinning Figure~\ref{fig:glitch_sig}. From this figure it can be seen that glitches, occurring in a pulsar which exhibits red noise, are unambiguously detectable, using all SKA-Low stations within 10 km of the core, when their amplitudes exceed $\Delta \nu > 10^{-9}\,$Hz. While these simulations reflect a somewhat conservative scenario, i.e., by using a comparatively faint ($1.1\,$mJy) pulsar, they highlight that small glitches are still detectable with high confidence. For brighter pulsars, which are expected to yield smaller \acs{ToA} uncertainties, glitches of significantly lower amplitude would be detectable with even greater significance using SKA-Low. Such sensitive detections of glitches are crucial for disambiguating the effects of timing noise from glitches in pulsar timing residuals (e.g., \citealt{espinoza:2014mnras}) and providing constrains on the glitch mechanism. 

For these simulations we have demonstrated the expected glitch detection performance of the SKA using a single implementation of a specific detection algorithm (Gaussian Process Regression). While alternative approaches exist---some of which may employ different metrics or statistical frameworks to assess detection significance---the broader conclusions we draw should not be strongly dependent on the choice of algorithm. Fundamentally, all such methods aim to identify sharp, anomalous deviations in timing residuals indicative of a glitch. As such, the differences in performance we find between observing configurations are expected to hold across a range of reasonable detection strategies.

It is important to note that the online glitch detection method described here bases its decision solely on the most recently acquired \acs{ToA}, assessing whether a glitch is likely to have occurred given the preceding data. In principle, this approach could be used to generate real-time alerts when a \acs{ToA} arrives significantly early or late relative to a prediction derived from prior \acsp{ToA}. However, any glitch detection strategy that relies on historical data must carefully consider both the observational cadence and the quantity of data used to construct predictive models. If the considered pre-glitch interval is too long, the effects of red noise must be taken into account, which can obscure smaller glitches. Conversely, using too few \acsp{ToA} can result in poor predictive power, particularly at low cadence. High-cadence observations also pose challenges: if the pulsar has not yet accumulated sufficient phase offset by the first post-glitch epoch, the event may not trigger an immediate alert. Nevertheless, such glitches—along with those falling below the real-time sensitivity limits discussed earlier—may still be identifiable retrospectively, as the discontinuous phase change becomes evident with additional post-glitch \acsp{ToA}. For this reason, regular offline glitch searches (e.g., \citealt{espinoza:2014mnras}) remain essential for compiling the glitch sample that is as complete as possible.

In order to maximise the efficient use of telescope time and avoid diverting follow-up resources unnecessarily, it is essential that any glitch detection and alert systems consider not only timing residuals, but also changes in the pulse profile. Events such as mode switching, intermittency (leading to non-detections), or radio-frequency interference can all introduce discontinuities into \acsp{ToA}. Robust classification frameworks that jointly evaluate timing anomalies and pulse shape changes are therefore critical to distinguishing true glitches from other forms of variability, and ensuring that real-time alerts are both sensitive and reliable.


\subsection{Detecting free precession with the SKAO}

Free precession introduces a periodic modulation in pulsar timing residuals (see Section~\ref{sec:precession}), with a sinusoidal morphology superimposed upon the otherwise smooth spin-down evolution of the star. The period $P_{\rm mod}$ of this modulation may span timescales up to several years, depending on the \ac{NS}’s internal structure and ellipticity. These amplitude variations are, in principle, detectable provided they introduce residual fluctuations that exceed the average \acs{ToA} uncertainty. Moreover, the changing orientation of the pulsar beam with respect to the line of sight, caused by precession, may lead to periodic variations in the observed pulse profile shape, introducing additional phase shifts in the timing residuals, and also in the observed polarisation position angle (e.g., \citealt{gao_etal_23, Desvignes+2024, Basu:2024qwt}), which can provide constraints on the pulsar's geometric asymmetry. 

\begin{figure}[b!]
\centering
\includegraphics[width=0.53\textwidth]{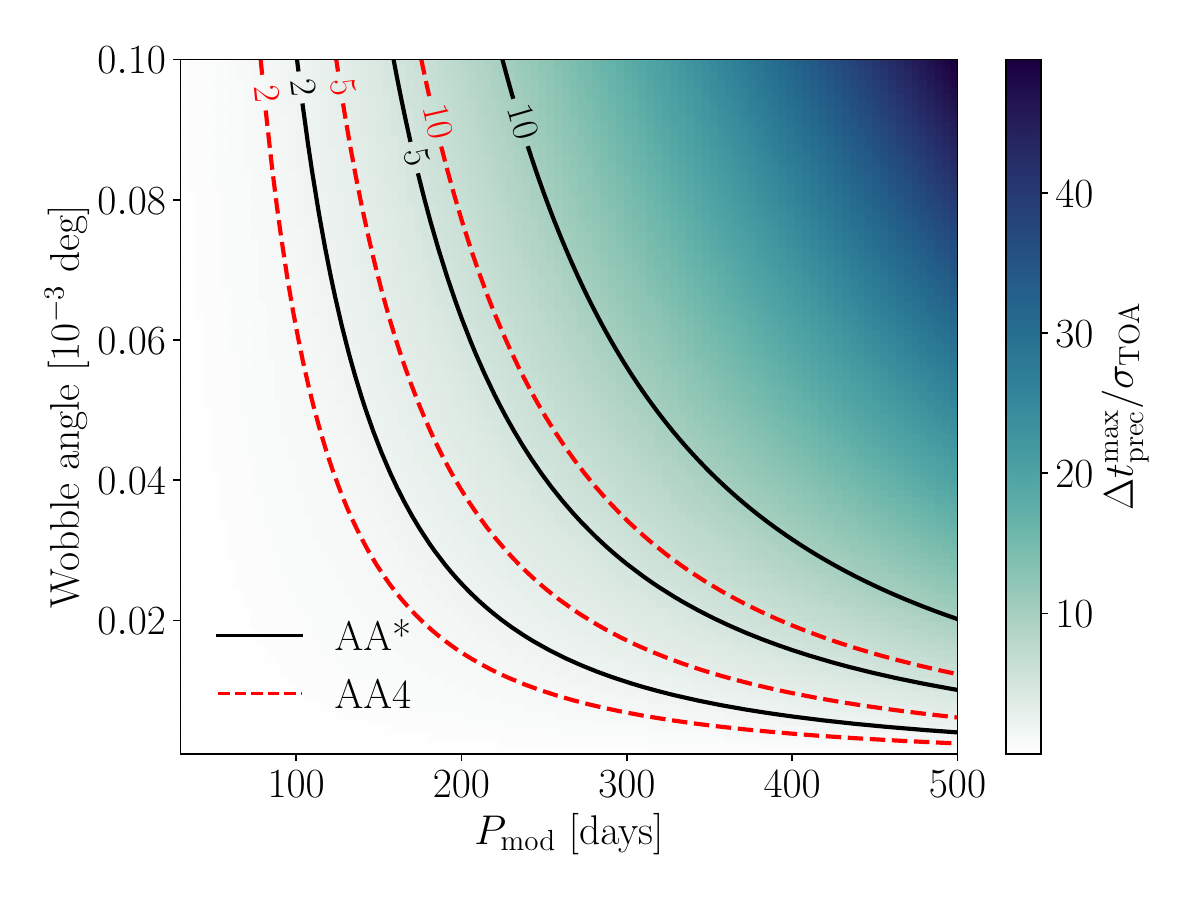}
\caption{Contour map showing the significance of the effects of free precession on pulsar timing residuals for a range of wobble angles and precession periods for SKA-Low during AA* (black solid line) and AA4 (red dashed line). Darker regions indicate parameter combinations where the influence of precession exceeds the \acs{ToA} uncertainty, signifying a higher likelihood of detection. See Section~\ref{case_study} for the pulsar parameters used for this simulation.}
\label{fig:ska_precession}
\end{figure}

To examine the ability of the SKA to resolve free precession in pulsar timing data, we utilise timing residuals generated as part of the simulations described in Section~\ref{sec:glitchobs}, for each array assembly, assuming each \acs{ToA} was measured using all available SKA-Low stations, as for a given pulsar, a higher timing precision is achievable. For simplicity, we did not include the effects of red noise, or any glitches in this analysis. We then inject oscillations introduced by free precession into the residuals. These oscillations are given by Equation~\eqref{eq:prece} and account for a range of wobble angles, $\theta$, and precession periods, $P_{\rm mod}$. Adapted from Equation (61) of \cite{Jones2001}, Equation~\eqref{eq:prece} describes the timing residuals expected from a precessing pulsar acted upon by the torque from the magnetosphere:
\begin{equation}\label{eq:prece}
    \Delta t_{\rm prec} = \frac{\cot{\chi} }{2\pi^2} \left ( \frac{P^2 \,\theta }{\tau_{\rm e} \epsilon_{\rm eff}^2 } \right ) \sin{(\dot \psi t + \psi_0)}, 
\end{equation}
where, $\epsilon_{\rm eff} = P/P_{\rm fp}$, $\dot \psi = 2\pi/P_{\rm fp}$ and $\tau_{\rm e} = P/\dot P$, where $P$ and $\dot P$ are the pulsar's spin period and derivative.

We then compare the peak-to-peak amplitude of the resulting timing residuals described by Equation~\eqref{eq:prece} to $\sigma_{\rm ToA}$, the ratio of which forms a measure of significance that we use to generate sensitivity curves as a function of $P_{\rm mod}$ and the wobble angle $\theta$. These are shown in Figure~\ref{fig:ska_precession}. The black and red lines show contour intervals corresponding to the 2, 5, and 10$\sigma$ levels of detection significance for AA* and AA4, respectively. These curves demonstrate that, assuming timing noise is negligible, AA4 will enable the detection of free-precession with over 5$\sigma$ confidence even for extremely small wobble angles ($\sim 2 \times 10^{-5}$ degrees) and precession periods of around 150 days. 


\subsection{Synergies with multiwavelength and multi-messenger facilities}
\label{sec:synergies}

Above, we have outlined how mass, spin and moment of inertia measurements using radio observations with the SKA will constrain the dense matter \ac{EoS}. These measurements are complemented by information from other techniques, in particular using X-ray and \ac{GW} data.

The former relies on inferring mass and radius via the technique of \acf{PPM}. X-rays emitted from the hot magnetic polar caps of rotation-powered millisecond X-ray pulsars pick up the imprint of mass and radius, thanks to various relativistic effects. By modelling the emission and using relativistic ray-tracing, one can infer not only the mass and radius but also the properties of the polar caps~\citep[see][and references therein]{Bogdanov:2019qjb,Bogdanov:2021yip}. Using data from the NICER telescope~\citep{Gendreau16}, the technique has now yielded good constraints for three millisecond pulsars: PSR~J0030$+$0451~\citep{Riley:2019yda, Miller19,Vinciguerra:2023qxq}; PSR~J0740$+$6620~\citep{Salmi24a, Dittmann24}; PSR~J0437$-$4715~\citep{Choudhury:2024xbk}; and weak constraints on PSR~J1231$-$1411~\citep{Salmi24b}. Over the course of its remaining mission lifetime, NICER is expected to deliver larger datasets that will yield tighter constraints for these sources, in addition to measurements for at least three more sources. We are also now starting to apply the \ac{PPM} technique to the population of accreting \acp{NS}~\citep{Poutanen03,Kini24,Dorsman25,Salmi25}. Future telescopes such eXTP~\citep{Li:2025uaw} and NewAthena~\citep{Cruise:2024mgo}, launching in the 2030s, are expected to result in even better constraints.  

\begin{figure}
\centering
\includegraphics[width=0.5\textwidth]{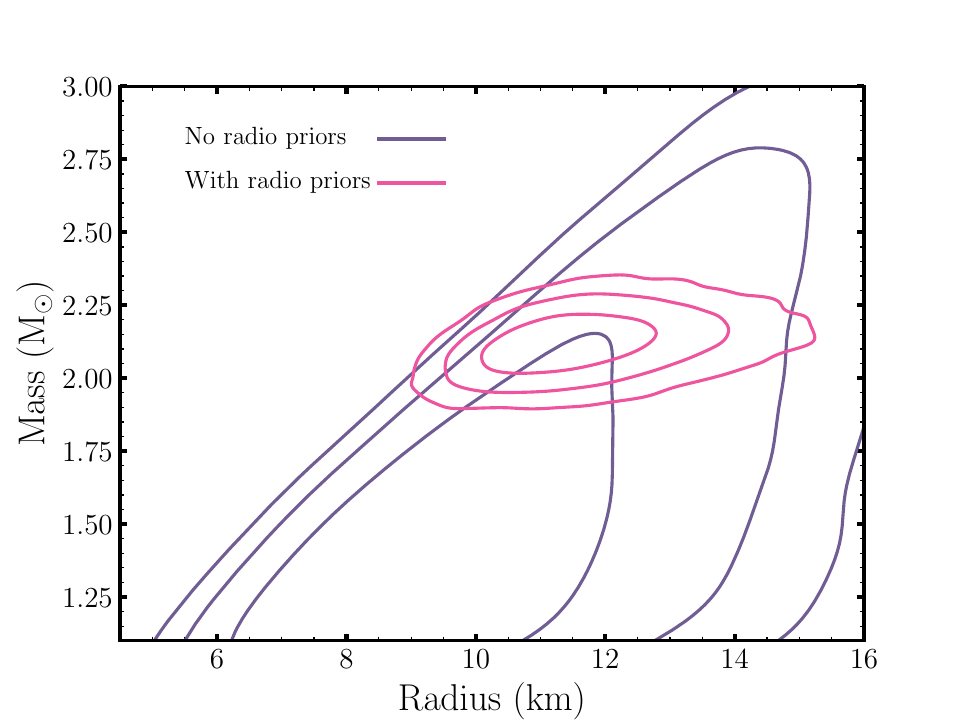}
\caption{This figure, adapted from Figure 5 of~\citet{Riley:2021pdl}, illustrates the impact of the radio timing information (mass, distance and inclination) on the mass, $M$, and equatorial radius, $R$, inferred from pulse profile modelling of NICER data of the high-mass pulsar PSR~J0740$+$6620. It shows the two-dimensional marginal PDFs conditional on the informative radio priors from~\citet{Fonseca:2021wxt} (pink) and a diffuse uninformative prior in the absence of radio data (purple). The contours are the credible regions containing 68.3\% 95.4\% and 99.7\% of the posterior mass.  Without the radio timing information, the inferred mass-radius constraints are too broad to be useful for \ac{EoS} analysis.}
\label{fig:PPMnomass}
\end{figure}

In principle, the results from \ac{PPM} provide an independent check on the \ac{EoS} constraints delivered by SKA measurements but in practice the relationship is much closer. The NICER millisecond pulsars are also radio pulsars, and where available the radio-derived mass measurement (along with distance and inclination) is used as a prior in the \ac{PPM}. This was the case for PSR~J0740$+$6620~\citep{Fonseca:2021wxt} and PSR~J0437$-$4715~\citep{Reardon24}: In both cases, the mass measurement played a critical role in enabling the inference as illustrated in Figure~\ref{fig:PPMnomass}. Improved mass priors provided by the SKAO for key \ac{PPM} sources will therefore be vital, highlighting the need for flexibility in timing those sources that are most promising for \ac{PPM} analysis. Information about the magnetospheric configuration (such as the magnetic and observer inclination angles;~\citealt{Oswald2025_MAG}), derived from radio observations with the SKA, is also expected to help reduce uncertainties relating to geometric/polar cap priors and model space involved in \ac{PPM}~\citep[see][for an example of a source where there are two potential mass-radius solutions associated with different magnetospheric geometries]{Vinciguerra:2023qxq}, resulting in tighter and more robust mass-radius inferences.

Looking further ahead, the feasibility of measuring the \ac{NS} radius of newly discovered pulsars by the SKA with X-ray \ac{PPM} not only depends strongly on the precise knowledge of the priors  outlined above, but also on the X-ray flux of the source (usually ranging between $10^{-13}-10^{-14}$\,erg~s$^{-1}$~cm$^{-2}$), the complexity of its surface temperature distribution, and the effective area and background knowledge of the available X-ray instrument. Consequently, the large range of parameters involved does not allow a firm identification of a distance limit below which we will be able to infer \ac{NS} radii for new SKA sources. However, especially with the future availability of NewAthena, X-ray observations significantly shorter than what is currently necessary with NICER will enable \ac{EoS} constraints (with a $<$3\% accuracy in the $M$-$R$ plane) for objects as distant as 1.5\,kpc~\citep{Cruise:2024mgo}.

Collectively, electromagnetic constraints provide an excellent counterpart to \ac{GW} observations and the opportunity to test underlying systematics for \ac{EoS} inferences. The potential of this synergy was exemplified by the observation of \acp{GW} from the binary \ac{NS} merger GW170817 by the LIGO-Virgo-KAGRA collaboration~\citep{LIGOScientific:2017vwq} and its corresponding electromagnetic counterpart~\citep{LIGOScientific:2017ync}. This event specifically highlighted how \acp{GW} constrain the \ac{NS} \ac{EoS} through measurements of the tidal deformation during inspiral~\citep{Chatziioannou:2024tjq}, ruling out the stiffest \acp{EoS}~\citep{LIGOScientific:2018hze}. Combining measurements of the \ac{NS}'s tidal properties with universal relations that connect the tidal parameters with $I$ and the \ac{NS}'s quadrupole moment might also allow new constraints on the latter two quantities~\citep{Yagi:2013awa}. The increased sensitivity of third-generation facilities such as the Einstein Telescope (ET)~\citep{Abac:2025saz} or Cosmic Explorer~\citep{Evans:2021gyd} will enable multiple binary NS merger detections, providing a wealth of new information on the dense matter \ac{EoS}. Combined with electromagnetic measurements, \ac{GW} detections from binary NS mergers might ultimately allow us to set tighter constraints on the possible existence of more exotic phases of matter in \ac{NS} cores, such as anisotropic states~\citep{Zuraiq:2023bpw} or twin stars~\citep{Christian:2021uhd}.

Complementary to terrestrial \ac{GW} observatories, the Laser Interferometer Space Antenna (LISA) is also expected to detect \acp{GW} emitted by a small population of `ultra-compact' ($P_{\rm b} \lesssim 10$\,min) relativistic double NS systems, that are otherwise undiscoverable by traditional radio pulsar surveys. A fraction of these pulsars will likely be detectable by the SKA through the use of LISA-informed radio searches \citep{Kyutoku:2019MNRAS}. Timing observations of these systems with the SKAO telescopes would enable precision measurements of Lense-Thirring precession, providing additional independent constraints on the \ac{NS} moment of inertia, thus further informing the \ac{EoS}~\citep{Thrane:2020MNRAS}.

In addition to detecting merger signals and ultra-compact binaries, future \ac{GW} interferometers may also provide an additional window into \ac{NS} interiors. As discussed before, a \ac{NS} is expected to not be spherical, but rather posses an ellipticity $\epsilon$. If this deformation, or `mountain', is not axisymmetric, it is swept around by rotation and sources continuous \ac{GW} emission at a frequency $\nu_{\rm gw}$, at either the rotation frequency $\nu_s$, such that $\nu_{\rm gw}=\nu_s$, or twice the rotation frequency, i.e., $\nu_{\rm gw}=2\nu_s$~\citep{Bonazzola96, Jaranowski98, Gittins24}. Microphysical modelling of the \ac{NS} crust suggests that $\epsilon\lesssim 10^{-6}$~\citep{Jones2001, Johnson2013, Gittins21a, Gittins21b} (although some \acp{EoS} predict exotic phases in the core that can sustain shearing and support larger ellipticities, \citealt{Haskell07}; larger deformations might also exist in superconducting \ac{NS} interiors,~\citealt{Das2025}), while astrophysical modelling of the evolution of millisecond pulsars in the $P$-$\dot{P}$ plane is consistent with a residual ellipticity in old \acp{NS} (possibly sourced by a buried magnetic field in the superconducting core) of $\epsilon\approx 10^{-9}$~\citep{Woan18}. These signals have not yet been detected, but current searches in LIGO-Virgo-KAGRA data are beginning to investigate astrophysically significant parameter spaces~\citep{HaskellBejger23}. 

\begin{figure}
\centering
\includegraphics[width=0.48\textwidth]{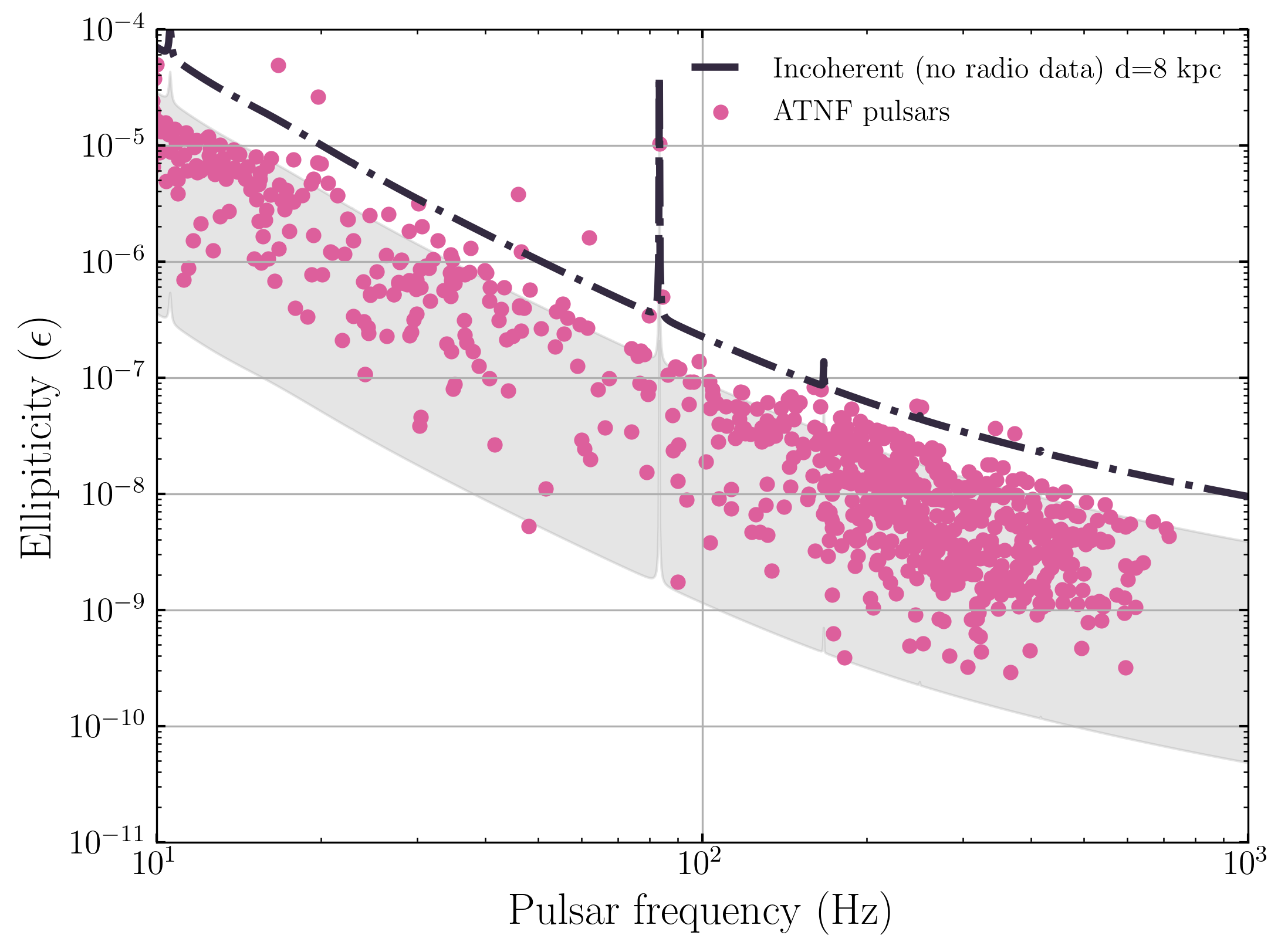}
\caption{Minimum detectable ellipticity for continuous \ac{GW} searches by ET. The black curve assumes a blind search with no radio data (i.e., an incoherent search with a total duration of 1 year with 10 days coherence time, as described in~\citealt{Branchesi:2023mws}) and an ellipticity limit calculated assuming a source at $8\,$kpc. The dots represent the minimum ellipticity detectable for pulsars in the ATNF Pulsar Catalogue~\citep[][\href{https://www.atnf.csiro.au/research/pulsar/psrcat/}{https://www.atnf.csiro.au/research/pulsar/psrcat/}]{manchester:2005atnf}, assuming in this case that radio ephemeris enable a full 1 year coherent integration. The grey band highlights the region between $0.1\,$kpc and $8\,$kpc. Note that we plot ellipticity versus the rotation frequency of the pulsar itself, not the GW frequency, which will be twice the stars' rotation frequency.}
\label{fig:ETpulsars}
\end{figure}

However, the true revolution will come with the next generation of ground-based detectors. In particular, ET is expected to see continuous \acp{GW} from several hundreds of known pulsars if their ellipticities are at the higher end of the expected range, and a few tens of pulsars if these are closer to the lower limit~\citep{Abac:2025saz}. Note that these numbers are based on \textit{known} pulsars only, i.e., those for which ephemeris are available from electromagnetic observations, typically in radio. This allows us to constrain parameters in the \ac{GW} search pipelines, and search coherently over long stretches of data. If, on the other hand, no ephemeris are available, one must search over a much larger parameter space, essentially performing a blind search, making a coherent search computationally impossible and leading to a much lower sensitivity~\citep{Riles23, Wette23}. This key difference is illustrated in Figure~\ref{fig:ETpulsars}. In practice, this means that blind searches are unlikely to reveal signals from small ellipticity NSs, unless they happen to be very close~\citep{Dergachev20}, probing closer to $\epsilon\approx 10^{-7}$ for the bulk of the population of rapidly rotating NSs in the galaxy~\citep{Branchesi:2023mws}. However, as the SKA is expected to find many more pulsars~\citep{Keane2025_SKA_Census, Levin2025_SKA_NSpop}, this will drastically increase the possibilities of detecting continuous \acp{GW} with third-generation \ac{GW} detectors, and maximise the possibility of studying signals close to the lower limit of theoretical predictions. This is not feasible without radio ephemeris, and will allow us to investigate aspects of the high density \ac{EoS} that are complementary to those probed by compact binary coalescences~\citep{Jones25}.

\section{Conclusions}
\label{sec:conclusions}

\acp{NS} offer access to the fundamental properties of matter at high densities, low temperatures, and large proton-neutron asymmetries---conditions that cannot be recreated on Earth. Their extreme environments make \acp{NS} unique laboratories for probing the nuclear \ac{EoS} and dense matter superfluidity. High-precision radio pulsar timing is a cornerstone of this endeavour, allowing us to infer global stellar properties like mass, moment of inertia, and spin frequency. When combined with observations in the X-rays, which measure the stellar radius, pulsar timing yields powerful constraints on the dense matter \ac{EoS}. Beyond global properties, pulsar timing also grants access to local phenomena, particularly glitches. These sudden spin-ups, now observed across many young pulsars, are the only observational window we currently have to study the dynamical properties of \ac{NS} superfluids.

The SKA, especially once AA4 comes online, will be transformative for this field. Its unprecedented sensitivity and wide survey capabilities will allow us to time existing \ac{NS} systems with even higher precision and greatly expand the population of known pulsars, including new highly relativistic binaries and fast rotators. This will enable novel high-precision $M$ measurements and $I$ constraints with single-digit uncertainties, potentially pushing the limits of existing \ac{EoS} models. Future SKA observations will also open up a new regime in glitch science and potentially the detection of free precession, enabling systematic studies of superfluid dynamics across a diverse pulsar population.

To achieve these goals, the SKA must operate in a range of observing modes. Large-scale surveys with SKA-Low and SKA-Mid will be essential for uncovering new systems with extreme and \ac{EoS}-relevant properties. Following these sources up with SKA-Mid for several hours with monthly cadence will be essential to build up the baselines relevant for the kinds of \ac{EoS} constraints outlined in this paper. To study glitches and other rotational irregularities, continuous monitoring over a wide range of timescales---from seconds to years---will be required. Given the transient nature of glitches and precession, these science cases also demand operational flexibility, including subarraying capabilities and a commensal timing mode, to capture and characterise events in real time.

Finally, synergies with other facilities, particularly X-ray observatories like NICER, and the upcoming NewAthena, and next-generation \ac{GW} facilities such as ET, will be essential to constraining subatomic physics. These multi-wavelength and multi-messenger approaches will not only provide deeper insights into the properties of ultra-dense matter but also open the door to probing physics beyond the Standard Model, from \ac{DM} interactions to alternative theories of gravity.


\section*{Acknowledgments}
The authors would like to thank Jim Palfreyman for providing data of the 2016 Vela glitch for Figure~\ref{fig:crabvelaf0f1}.
Pulsar research at Jodrell Bank is supported by a consolidated grant (ST/T000414/1 and ST/X001229/1) from the UK Science and Technology Facilities Council (STFC).
V.~G. is supported by a UKRI Future Leaders Fellowship (grant number MR/Y018257/1). 
M.~E.~L. is supported by an Australian Research Council (ARC) Discovery Early Career Research Award DE250100508. 
D.~A. acknowledges support from an EPSRC/STFC fellowship (EP/T017325/1). 
P.~C. acknowledges the support from the European Union's HORIZON MSCA-2022-PF-01-01 Programme under Grant Agreement No.~101109652, project ProMatEx-NS. This project has received funding from the European Union’s Horizon 2020 research and innovation programme under the Marie Skłodowska-Curie grant agreement No.~101034371.
D.~I.~J acknowledges support from the STFC via grant No.~ST/R00045X/1.
M.~O. is supported by the Agence Nationale de la Recherche (ANR) under contract ANR-22-CE31-0001-01.
N.~R. is supported by the European Research Council (ERC CoG No.~817661 and ERC PoC No.~101189496), and grants SGR2021-01269, ID2023-153099NA-I00, and CEX2020-001058-M.
V.~S. gratefully acknowledges support from the UKRI-funded ``The next-generation gravitational-wave observatory network'' project (Grant No. ST/Y004248/1) and the support by Funda\c c\~ao para a Ci\^encia e Tecnologia (FCT), project 2023.10526.CPCA.A2 with DOI identifier 10.54499/2023.10526.CPCA.A2.
A.~L.~W. acknowledges support from NWO grant ENW-XL OCENW.XL21.XL21.038 and ERC Consolidator grant No.~865768 AEONS.


\bibliographystyle{apsrev4-1}
\bibliographystyle{aasjournal}
\bibliography{bibliography_OJA}

\end{document}